\documentclass[12pt]{article}
\usepackage{amsmath}
\usepackage{graphicx}
\usepackage{enumerate}
\usepackage{url} 

\newcommand{\blind}{1}

\usepackage[authoryear]{natbib}
\usepackage{amsthm}
\RequirePackage[colorlinks,citecolor=blue,urlcolor=blue]{hyperref}
\usepackage{amssymb}

\usepackage{amsfonts}
\usepackage{comment}
\usepackage{mathrsfs}
\usepackage{color}
\usepackage{pgfplots}
\usepackage{filecontents}
\usepackage{subfigure}
\usepackage{paralist}
\usepackage{array}
\usepackage{amsmath}
\usepackage{bm}
\usepackage{rotating}
\usepackage{booktabs}
\usepackage{multirow}

\usepackage{algorithm}
\usepackage{algorithmic}

\graphicspath{{figs_new/}}  
\usepackage{subfigure} 

\allowdisplaybreaks[4]

\theoremstyle{plain}

\newtheorem{theorem}{Theorem}[section]
\newtheorem{lemma}[theorem]{Lemma}

\theoremstyle{remark}

\newcommand{\CUT}[1]{}

\begin{document}

\def\spacingset#1{\renewcommand{\baselinestretch}%
{#1}\small\normalsize} \spacingset{1}


\if1\blind
{
  \title{\bf Optimized and regularly repeated lattice-based Latin hypercube designs for large-scale computer experiments}
  \author{Xu He\\
    State Key Laboratory of Mathematical Sciences (SKLMS), \\Academy of Mathematics and Systems Science, \\Chinese Academy of Sciences, hexu@amss.ac.cn, \\
Junpeng Gong\\
    School of Mathematical Sciences, University of Chinese Academy of \\Sciences and State Key Laboratory of Mathematical Sciences (SKLMS), \\Academy of Mathematics and Systems Science, \\Chinese Academy of Sciences,  gongjunpeng21@mails.ucas.ac.cn\\
and\\
Zhaohui Li\\
    State Key Laboratory of Mathematical Sciences (SKLMS), \\Academy of Mathematics and Systems Science, \\Chinese Academy of Sciences, lizh@amss.ac.cn
}
  \maketitle
} 
\fi

\if0\blind
{
  \bigskip
  \bigskip
  \bigskip
  \begin{center}
    {\LARGE\bf Optimized and regularly repeated lattice-based Latin hypercube designs for large-scale computer experiments}
\end{center}
  \medskip
} \fi

\bigskip
\begin{abstract}
Computer simulations serve as powerful tools for scientists and engineers to gain insights into complex systems. Less costly than physical experiments, computer experiments sometimes involve large number of trials. Conventional design optimization and model fitting methods for computer experiments are inefficient for large-scale problems. In this paper, we propose new methods to optimize good lattice point sets, using less computation to construct designs with enhanced space-filling properties such as high separation distance, low discrepancy, and high separation distance on projections. These designs show promising performance in uncertainty quantification as well as physics-informed neural networks. We also propose a new type of space-filling design called regularly repeated lattice-based Latin hypercube designs, which contain lots of local space-filling Latin hypercube designs as subdesigns. Such designs facilitate rapid fitting of multiple local Gaussian process models in a moving window type of modeling approach and thus are useful for large-scale emulation problems. 
\end{abstract}

\noindent%
\vfill

\newpage
\spacingset{1.9} 
\section{Introduction}
\label{sec:intro}

Computer simulations serve as powerful tools for scientists and engineers, enabling them to gain deeper insights into complex systems~\citep{Santner:book,Gramacy:book}.
Less costly than physical experiments, computer experiments sometimes involve large number of trials.
The large scale introduces significant computational challenges, particularly in the construction of experimental designs and the subsequent model fitting process.
This paper introduces novel experimental design methods aimed at overcoming these challenges.

Latin hypercube designs (LHD)~\citep{McKay:1979} are the most widely used type of experimental design for computer experiments.
They are particularly appealing when input factors exhibit sparsity, as they uniformly cover the domain of each individual factor~\citep{Morris:1995}.
LHDs with additional space-filling properties, such as low discrepancy~\citep{Fang:book}, high separation distance, or low fill distance~\citep{Johnson:1990}, are more desirable than random LHDs when most or all inputs are influential.

There are three primary approaches for generating space-filling LHDs.
The first approach involves algebraic construction methods, which have been proposed by \citet{Tang:1993}, \citet{He:2013}, \citet{Xiao:2017}, \citet{Xiao:Xu:2018}, \citet{Zhou:2020}, and others.
These methods are computationally efficient and thus suitable for constructing large-scale designs.
However, they impose strict constraints on the sample size $n$ and the number of factors $d$.
As a result, a more widely used approach is to numerically optimize specific space-filling properties of LHDs~\citep{Morris:1995}.
However, numerical optimization algorithms are computationally inefficient for constructing large-scale designs due to the high-dimensional nature of the optimization problem, which contains numerous local optima.
Given finite computing resources, it is inherently difficult to find an optimal or near-optimal solution.

The third approach restricts the search to good lattice point sets (GLP)~\citep{Korobov:1959}, which are a specific type of LHD with a lattice structure.
The lattice constraint offers two key advantages: 
First, the lattice structure is beneficial for uncertainty quantification, particularly in estimating the distribution of outputs when some inputs or model parameters are random with known prior distributions~\citep{Hua:Wang:book,Sloan:Joe:book}. 
Moreover, imposing the lattice constraint significantly reduces the search space, making it numerically easier to identify a high-quality design.
Despite these advantages, for large values of $n$, there remain many distinct lattice point sets, making it challenging to identify the optimal set.
\citet{Korobov:1959} recommended an exhaustive search over all power generators, as lattice point sets generated from this type of generators typically exhibit low discrepancy.
However, searching over too few options increases the likelihood of missing the optimal design.
Recently, \citet{Vazquez:2024} proposed an integer programming algorithm aimed at maximizing the separation distance of GLPs. 
However, this method focuses on scenarios where $n \leq 113$. 
Finally, algebraic methods to construct GLPs with excellent separation distance or discrepancy have been proposed by \citet{Zhou:2015}, \citet{Wang:Xiao:Xu:2018}, \citet{Sun:Wang:Xu:2019}, \citet{Wang:Sun:Xu:2020}, among others. 
However, similar to algebraic constructions for non-lattice LHDs, these methods are applicable only for specific values of $n$ and $d$.

In this work, we introduce new shortcut formulas for computing separation distance, discrepancy, and projection properties of GLPs. 
By reducing the computational complexity from $O(n^2)$ to $O(n)$ or $O\{\log(n)\}$, these formulas significantly improve the scalability of the approach, enabling its application in large-scale settings. 
When combined with an efficient optimization algorithm, the generated designs demonstrate superior space-filling properties while requiring less computational effort.
When applied to the classic problem of estimating the mean simulation outcome, these designs consistently outperform others for $n \geq 100$.
Owing to this advantage, they are useful in training physics-informed neural networks (PINNs)~\citep{raissi2019physics}, which typically requires a large set of design points. 
The PINN employs a neural network to approximate the solution of a partial differential equation (PDE). 
To minimize the integrated squared residuals, PINNs are trained to minimize the sum of squared residuals at a high-volumn of design points covering the input space. 
Consequently, designs that minimize numerical integration error are ideal for training PINNs~\citep{wu2023comprehensive}. 
In particular, \citet{matsubara2023good} has proposed a method to construct lattice-based Latin hypercube designs and shown they perform well in PINNs. 
Our numerical results show that designs constructed from our proposed algorithms result in much lower training errors than existing methods, including those proposed in \citet{matsubara2023good} across two PINN problems. 

Our second major contribution is the proposal of a new type of design, termed regularly repeated lattice-based Latin hypercube designs (RLHD), and the demonstration of how to use them for large-scale emulation problems.
Gaussian process models (GPM) are commonly used for fitting surrogate models of computer simulations~\citep{Sacks:1989}.
When the simulation output exhibits sharp changes, a large number of simulation trials may be required to fit an accurate model.
However, fitting a GPM with $n$ samples involves inverting an $n\times n$ correlation matrix, which requires $O(n^3)$ operations and becomes impractical for large $n$.
Several approaches exist to accelerate the computation.
First, covariance tapering techniques that induce sparsity in the correlation matrix~\citep{Kaufman:2011} or low-rank approximation methods that reduce its rank~\citep{Cressie:2008} can make the matrix easier to invert.
However, this often results in reduced accuracy.
Second, rapid fitting can be achieved by using sparse grid designs in conjunction with specialized modeling techniques~\citep{Plumlee:2014,Ding:2022}.
However, sparse grids typically do not yield the most accurate surrogate models.
Third, fitting GPMs with subsamples is a reasonable approach, especially for simulations with sharp changes, where predictions are largely based on trials whose inputs are nearby~\citep{Kim:2005}.
A naive strategy involves dividing the input domain into $s$ subdomains and fitting separate local models.
Since each subdomain contains approximately $n/s$ points, the total computational cost for fitting all $s$ models is $O\{(n/s)^3s\} = O(n^3s^{-2})$, which is substantially lower than $O(n^3)$.
However, this approach may yield poor predictions for positions near the boundary of the subdomain to which they belong.
To address this issue, one can use a mixture of local models~\citep{Gramacy:2008,Ding:2011} or fit a separate local model for each testing location~\citep{laGP}.
However, for large-scale emulation problems, these strategies tend to introduce excessive additional computations. 

The RLHDs we propose are designs which contain a large amount of identical local Latin hypercube subdesigns. 
For each testing input, we can use the local model fitted by the surrounding local LHD to predict the outcome.
Since the subdesigns differ only by translation, the correlation matrices of the local models remain the same, assuming a stationary correlation function.
Therefore, fitting all local models requires $O(m^3)$ operations, the same as fitting a single local model, where $m$ is the subsample size.
As a result, making large-scale predictions using $m$-point subsamples from an RLHD is considerably faster than using $m$-point subsamples from other strategies.
More importantly, since all subsamples are LHDs with optimized space-filling properties, the local models are typically more accurate than those fitted using other strategies.
Since our method makes no assumption about the correlation kernel other than stationarity, it can be applied to stochastic computer experiments and combined with tapering or low-rank approximation methods.

The remainder of the paper is organized as follows.
First, in Section~\ref{sec:LHD}, we review methods for optimizing LHDs. 
In Section~\ref{sec:method}, we propose new methods for optimizing LHDs and regularly repeated designs. 
Numerical comparisons of space-filling properties, uncertainty quantification accuracy, PINN training accuracy, and large-scale emulation accuracy are presented in Section~\ref{sec:comp}. 
Final remarks are given in Section~\ref{sec:conc}. 
Proofs of theorems, an R package for efficiently generating our proposed designs, and the computer code used to produce the numerical results are provided in the supplementary material.

\section{Review of methods for optimizing LHDs}
\label{sec:LHD}

Throughout this paper, we denote the $(i,k)$-th entry of the matrix $\textbf{X}$ by $x_{i,k}$ and express the $i$-th row of $\textbf{X}$ as $\textbf{x}_i= (x_{i,1}, \ldots, x_{i,d})$.
An $n\times d$ matrix is called a centered Latin hypercube design (LHD) in $[0,1]^d$ if each of its columns is a permutation of the set $\{1/(2n),3/(2n),\ldots,(2n-1)/(2n)\}$. 
Sometimes we also represent an $n$-point design as a set containing $n$ elements. 
Algorithm~\ref{alg:LHD} provides the steps to generate a random LHD and 
Algorithm~\ref{alg:OLHD} describes a typical optimization procedure for constructing optimized LHDs (OLHDs), whose parameter values can be $q = 0.95$, $r = 10$, and $T = 2000$. 

\begin{algorithm}[!t]
\caption{Construction of random Latin hypercube designs (LHD)}\label{alg:LHD}
\begin{algorithmic}[1]
\STATE \textbf{Input:} sample size $n$, number of covariates $d$.
\STATE Independently generate $\pi_1,\ldots,\pi_d$, $d$ uniform permutations on $\{1,\ldots,n\}$;
\FOR{$i = 1, \ldots, n$ and $k = 1, \ldots, d$}
        \STATE let $x_{i,k} \leftarrow \pi_k(i)/n-1/(2n)$\label{step:LHD:perturb};
\ENDFOR
\STATE \textbf{Return} the $n \times d$ matrix $\textbf{X}$. 
\end{algorithmic}
\vspace*{-2pt}
\end{algorithm}

\begin{algorithm}[!t]
\caption{Construction of optimized Latin hypercube designs (OLHD)}\label{alg:OLHD}
\begin{algorithmic}[1]
\STATE \textbf{Input:} sample size $n$, number of covariates $d$, number of iterations $T$, optimization criterion $c(\cdot)$, temperature dropping speed $q$, initial temperature $r$.
\STATE Generate $\textbf{X}$, an LHD using Algorithm~\ref{alg:LHD}, and let $\textbf{X}_{\text{Best}} \leftarrow \textbf{X}$; 
\FOR{$t$ from 1 to $T$} 
\STATE Sample $k$ uniformly from $\{1,\ldots,d\}$ and sample $i \neq j$ uniformly from  $\{1,\ldots,n\}$; 
\STATE Switching the $(i,k)$th and the $(j,k)$th entries of $\textbf{X}$ to obtain a new design  $\textbf{X}_{\text{Try}}$; 
\STATE Compute the optimization criterion for the new design, $c(\textbf{X}_{\text{Try}})$;
\STATE If $c(\textbf{X}_{\text{Try}}) < c(\textbf{X}_{\text{Best}})$, let $\textbf{X}_{\text{Best}} \leftarrow \textbf{X}_{\text{Try}}$ and update $c(\textbf{X}_{\text{Best}})$; 
\STATE With $\min[ e^{-\{c(\textbf{X}_{\text{Try}}) - c(\textbf{X})\} /(q^t r)}, 1]$ chance let $\textbf{X} \leftarrow \textbf{X}_{\text{Try}}$ and update $c(\textbf{X})$; 
\ENDFOR
\STATE \textbf{Return} the $n \times d$ matrix $\textbf{X}_{\text{Best}}$. 
\end{algorithmic}
\vspace*{-2pt}
\end{algorithm}

Common criteria $c(\cdot)$ for OLHD that are the lower the better include centered discrepancy (CD), wrap-around discrepancy (WD), reciprocal separation distance (RS), approximate reciprocal separation distance (AS), projective separation distance (PS), among others. 
\citet{phip} found that while random LHDs are statistically uniform in the sense that each design point marginally follows the uniform distribution in $[0,1]^d$, the OLHDs obtained by optimizing CD, RS, AS, or PS are not statistically uniform. 
Consider the RS-optimized design as an example. To maximize the pairwise distance between design points, it is clearly advantageous to spread the points toward the boundary facets of $[0,1]^d$, resulting in a non-uniform distribution of the points. 
To address this issue, \citet{phip} proposed using the wrap-around distance 
$w( \textbf{x}_i - \textbf{x}_j )$ 
to replace the Euclidean distance
$\|\textbf{x}_{i} - \textbf{x}_{j}\|$
in the design criteria, where $w(\textbf{z}) = \| \textbf{z} - \lceil \textbf{z} \rfloor \|$ 
and $\lceil z \rfloor$ denotes the rounding of $z$ to the nearest integer.
By replacing distance measures in RS, AS, PS, and CD, we obtain 
the wrap-around reciprocal of the square root of separation distance (WS),
    \(
    c_{\text{WS}}(\textbf{X}) = \max_{i \neq j} [ \{ \sum_{k=1}^d w( x_{i,k} - x_{j,k} )^2 \}^{-1/2} ],
    \)
the approximated wrap-around separation distance (WA),
    \(
    c_{\text{WA}}(\textbf{X}) = [ \sum_{i < j}\{ \sum_{k=1}^d w( x_{i,k} - x_{j,k} )^2 \}^{-50/2} ]^{1/50},
    \)
the wrap-around projective separation distance (WP),
    \(
    c_{\text{WP}}(\textbf{X}) =  \sum_{i < j} \{ \prod_{k=1}^d w(x_{i,k} - x_{j,k})^{-2} \} / \{n(n-1)/2\} ]^{1/d}, 
    \)
and the wrap-around discrepancy (WD), 
\[ c_{\text{WD}}(\textbf{X}) = \left[ \sum_{\emptyset \subsetneq \textbf{U} \subset \{1,\ldots,d\}} \int_{[0,1]^{2|\textbf{U}|}} \left\{ \left| \textbf{X} \cap \textbf{W}_{\textbf{U}}(z_k-y_k) \right| / n - \prod_{k \in \textbf{U}} r(z_k-y_k) \right\}^2 \prod_{k \in \textbf{U}} (dy_k dz_k) \right]^{1/2}, \]
respectively, 
where $\lfloor z \rfloor$ denotes the highest integer that is no greater than $z$, 
$r(z_k) = z_k - \lfloor z_k \rfloor$ gives the fractional part of $z_k$, and $\textbf{W}_{\textbf{U}}(z_k-y_k) = [y_k,z_k]$ if $z_k\geq y_k$ and $\textbf{W}_{\textbf{U}}(z_k-y_k) = [0,z_k] \cup [y_k,1]$ otherwise. 
The $c_{\text{WD}}(\textbf{X})$ can also be computed through~\citep{Hickernell:1998:chapter} 
\(
c_{\text{WD}}(\textbf{X})^2 = \sum_{i,j=1}^n \prod_{k=1}^d \left\{ 1.25 + w(x_{i,k}-x_{j,k}+.5)^2 \right\} /n^2 - (4/3)^d. 
\)
Designs that are optimal under the WD, WS, WA, and WP criteria are nearly optimal under the CD, RS, AS, and PS criteria, respectively.
Meanwhile, \citet{phip} proved that OLHDs obtained by optimizing WD, WS, WA, and WP are statistically uniform and demonstrated that these designs perform better than their non-uniform counterparts for uncertainty quantification. 

It is straightforward to derive that, for any of these eight criteria, it takes \(O(n^2 d)\) operations to compute a single \(c(\textbf{X})\). Therefore, the \(T\) iterations of Algorithm~\ref{alg:OLHD} require a total of \(O(T n^2 d)\) operations. However, updating \(c(\textbf{X}_{\text{Try}})\) based on \(c(\textbf{X})\) may require significantly fewer operations.
For example, \citet{Jin:2005} showed that  
\( 
c_{AS}(\textbf{X}_{\text{Try}})^{50} = c_{AS}(\textbf{X})^{50} 
 + \sum_{h \neq i, h \neq j} [ \{ ( x_{j,k} - x_{h,k} )^2 + \sum_{l \neq k} ( x_{i,l} - x_{h,l} )^2 \}^{-50/2} - \{ \sum_{l=1}^d ( x_{i,l} - x_{h,l} )^2 \}^{-50/2} ] 
 + \sum_{h \neq i, h \neq j} [ \{ ( x_{i,k} - x_{h,k} )^2 + \sum_{l \neq k} ( x_{j,l} - x_{h,l} )^2 \}^{-50/2} - \{ \sum_{l=1}^d ( x_{j,l} - x_{h,l} )^2 \}^{-50/2} ].
\)
It is evident that updating \(c_{AS}(\textbf{X})\) to \(c_{AS}(\textbf{X}_{\text{Try}})\) requires \(O(n d)\) operations. As a result, Algorithm~\ref{alg:OLHD} can be implemented with a total of \(O(T n d)\) operations.
Similar techniques can be applied to the CD, WD, PS, WA, and WP criteria, with the formulas for CD and WD provided in \citet{Fang:2006}. However, popular implementations such as R package DiceDesign~\citep{R:DiceDesign}, SLHD~\citep{R:SLHD}, and MaxPro~\citep{R:MaxPro} do not utilize these shortcut techniques, meaning that the full algorithm still requires \(O(T n^2 d)\) operations.
For this reason, we provide R code in the supplementary material that implements the shortcut formulas for CD, WD, AS, PS, WA, and WP.

Let $\textbf{1}_d$ denote the $d$-vector consisting of ones. If $\textbf{v} \in \mathbb{Z}^d$ and $\boldsymbol{\delta} \in \mathbb{R}^d$, the set 
\begin{equation}\label{eqn:LLHD}
 \textbf{L}(n,\textbf{v},\boldsymbol{\delta}) = \{ \textbf{z} + i\textbf{v}/n + \boldsymbol{\delta}/n + \textbf{1}_d/(2n) : \textbf{z} \in \mathbb{Z}^p, i \in \mathbb{Z} \} \cap [0,1]^d 
\end{equation}
is a good lattice point set type of LHD (LLHD). 
From Lemma~\ref{lem:LLHD} below~\citep{Fang:book}, the construction of optimal LLHD reduces to the problem of choosing a generator $\textbf{v} \in \{1,\ldots,n-1\}^d$ and a shift vector $\boldsymbol{\delta} \in \{0,n-1\}^d$. 

\begin{lemma}\label{lem:LLHD}
Suppose any entry of $\textbf{v}$ is coprime to $n$ and $\boldsymbol{\delta} \in \mathbb{Z}^d$. 
Then $\textbf{L}(n,\textbf{v},\boldsymbol{\delta})$ in \eqref{eqn:LLHD} is an $n$-point LHD in $[0,1]^d$ and
$ \textbf{L}(n,\textbf{v},\boldsymbol{\delta}) = \{ r\{ i\textbf{v}/n+\boldsymbol{\delta}/n+ \textbf{1}_d/(2n) \} : i = 0,\ldots,n-1 \}$. 
\end{lemma}

However, there are many possible choices for \( \textbf{v} \) for medium-to-large \( n \) and thus it is not feasible to exhaustively test all \( \textbf{v} \)'s to find the optimal one. 
For this reason, \citet{Korobov:1959} suggested trying only power generator vectors, which can be expressed as \( \textbf{v} = (1, g, g^2, \ldots, g^{d-1}) \) with an integer \( g \) that is coprime to \( n \). 
We refer to PLHD as designs obtained by numerically optimizing a space-filling criterion over GLPs with such $\textbf{v}$. 
Although the construction of PLHDs is very fast, we find that in most cases the PLHDs are not optimal because too few choices are searched for.

\section{Design construction methodology}
\label{sec:method}

\subsection{Construction of optimal good lattice point sets}

In this subsection, we propose our method to generate optimized LLHDs. 
We focus on criteria using the wrap-around distance because, as discussed in Section~\ref{sec:LHD}, they lead to statistically-uniform designs whereas other discussed criteria do not. 
Theorem~\ref{thm:LLHD:evaluate} below gives formulas to evaluate the WS, WA, WP, and WD of one LLHD using $O(nd)$ operations. 

\begin{theorem}\label{thm:LLHD:evaluate}
Suppose each entry of $\textbf{v}$ is coprime to $n$, $\boldsymbol{\delta} \in \mathbb{Z}^d$, and $\textbf{L}(n,\textbf{v},\boldsymbol{\delta})$ is an LLHD in \eqref{eqn:LLHD}. 
Then  
\begin{eqnarray*}
 c_{\text{WS}}\{\textbf{L}(n,\textbf{v},\boldsymbol{\delta})\}   &=& \min_{i=1}^{n-1} \left[ \left\{ \sum_{k=1}^d w( i\textbf{v}_k/n )^2 \right\}^{-1/2} \right], \\
 c_{\text{WA}}\{\textbf{L}(n,\textbf{v},\boldsymbol{\delta})\}   &=& \left[ \frac{n}{2} \sum_{i=1}^{n-1} \left\{ \sum_{k=1}^d w( i\textbf{v}_k/n )^2 \right\}^{-50/2} \right]^{1/50}, \\
 c_{\text{WP}}\{\textbf{L}(n,\textbf{v},\boldsymbol{\delta})\}   &=& \left[  \sum_{i=1}^{n-1} \left\{ \prod_{k=1}^d w( i\textbf{v}_k/n )^{-2} \right\} / (n-1) \right]^{1/d}, \\
 c_{\text{WD}}\{\textbf{L}(n,\textbf{v},\boldsymbol{\delta})\}^2 &=& \sum_{i=1}^n \prod_{k=1}^d \left\{ 1.25 + w( i\textbf{v}_k/n -1/2 )^2 \right\} /n - (4/3)^d. 
\end{eqnarray*}
\end{theorem}

Since the $O(nd)$ operations are significantly less than the $O(n^2 d)$ operations required to evaluate the WS, WA, WP, or WD for an ordinary LHD, restricting to LLHDs allows for more designs to be searched with less computational cost.
Theorem~\ref{thm:LLHD:evaluate} also verifies that for LLHD, the WS, WA, WP, and WD are shift-invariant, 
a feature that is desirable for a criterion to possess according to \citet{Fang:book}. 
Owing to this property, the design construction problem reduces from the optimization of both $(\textbf{v}, \boldsymbol{\delta})$ to the optimization of solely $\textbf{v}$, 
which further accelerates the computation.

Let $ \textbf{P}(n) $ denote the set of positive integers smaller than $ n/2 $ that are coprime to $ n $, and let $ p(n) $ denote the cardinality of $ \textbf{P}(n) $.
When $ p(n) \leq d $, we propose using Algorithm~\ref{alg:LLHD} below to generate LLHDs. 
In the algorithm, we force the entries of $ \textbf{v} $ to be distinct because when $ v_k = v_l $, the $ \textbf{L}(n, \textbf{v}, \boldsymbol{\delta}) $ will be poor after being projected onto the $ (k,l) $-th dimensions. 
On the other hand, when $ d > p(n) $, it is not possible to avoid $ v_k = v_l $ for some $ (k,l) $ pairs. In such cases, we propose to generate an LLHD with $ d - \lfloor d/p(n) \rfloor p(n) $ columns and then supplement it by adding $ \lfloor d/p(n) \rfloor $ many $ \textbf{L}\{n, \textbf{P}(n), \boldsymbol{\delta}\} $'s.

\begin{algorithm}[!t]
\caption{Construction of optimized LLHD}\label{alg:LLHD}
\begin{algorithmic}[1]
\STATE \textbf{Input:} sample size $n$, number of covariates $d$, number of iterations $T$, number of random starts $Q$, optimization criterion $c(\cdot)$.
\STATE Sample $\boldsymbol{\delta}$ from the uniform distribution on $\{0,\ldots,n-1\}^d$;
\STATE Initialize $c_{\text{best}} \leftarrow +\infty$;
\FOR{$q$ from 1 to $Q$} 
\STATE Randomly sample $v_1,\ldots,v_d$ from $\textbf{P}(n)$ without replacement; 
\STATE Compute $c\{\textbf{L}(n,\textbf{v},\boldsymbol{\delta})\}$;
\FOR{$t$ from 1 to $T/Q$} 
\STATE Sample $k$ from the uniform distribution on $\{1,\ldots,d\}$ and sample $\tilde v_k$ from the uniform distribution on $\textbf{P}(n) \setminus \{v_1,\ldots,v_d\}$;
\STATE Let $\tilde{\textbf{v}}$ be the vector whose $k$th entry is $\tilde v_k$ and other entries are the same to $\textbf{v}$; 
\STATE Compute the optimization criterion for the new design, $c\{\textbf{L}(n,\tilde{\textbf{v}},\boldsymbol{\delta})\}$;
\STATE If $c\{\textbf{L}(n,\tilde{\textbf{v}},\boldsymbol{\delta})\} \leq c\{\textbf{L}(n,\textbf{v},\boldsymbol{\delta})\}$, update $\textbf{v} \leftarrow \tilde{\textbf{v}}$; 
\ENDFOR
\STATE If $c\{\textbf{L}(n,\textbf{v},\boldsymbol{\delta})\} < c_{\text{best}}$, update $\textbf{v}_{\text{best}} \leftarrow \textbf{v}$ and $c_{\text{best}} \leftarrow c\{\textbf{L}(n,\textbf{v},\boldsymbol{\delta})\}$;
\ENDFOR
\STATE \textbf{Return} the $\textbf{v}_{\text{best}}$, $\boldsymbol{\delta}$, and $\textbf{L}(n,\textbf{v}_{\text{best}},\boldsymbol{\delta})$. 
\end{algorithmic}
\vspace*{-2pt}
\end{algorithm}

Algorithm~\ref{alg:LLHD} is a neighborhood search algorithm in which the neighbourhood consists of LLHDs that differs in one dimension. 
Clearly, each neighbourhood contains roughly $p(n)d \leq nd/2$ elements. 
In contrast, each neighbourhood for Alghorithm~\ref{alg:OLHD} contains roughly $n^2d/2$ elements. 
As a result, Algorithm~\ref{alg:LLHD} usually reaches to a local optimum with much fewer operations than in optimizing OLHDs. 
From numerical results, $5p(n)d$ iterations seem to be enough for reaching a local optimum. 
For this reason, we recommend to try $Q=\max[ \lfloor T/\{5p(n)d\} \rfloor, 1]$ random starts. 

Next, we propose two criteria that are even faster to evaluate.
Following the effect hierarchy principle, lower-order interactions are more important than higher-order interactions for space-filling designs~\citep{Shi:Xu:2023+}. 
Since all LHDs achieve optimal separation distance and discrepancy on univariate projections, criteria reflecting the bivariate projection properties of designs are reasonable for discriminating between LHDs. 
We thus consider the bivariate separation distance criterion and its wrap-around version, 
\[ c_{\text{RS2}}(\textbf{X}) = \sum_{k < l} c_{\text{RS}}(\textbf{X}_{k,l}), \quad c_{\text{WS2}}(\textbf{X}) = \sum_{k < l} c_{\text{WS}}(\textbf{X}_{k,l}), \]
where $\textbf{X}_{k,l}$ denote the design $\textbf{X}$ after projected to the $k$th and $l$th dimensions. 
Clearly, $c_{\text{WS2}}(\textbf{X}) \geq c_{\text{RS2}}(\textbf{X})$ for any design $\textbf{X}$. 
Theorem~\ref{thm:LLHD:WD2RD2} below shows that the two criteria are equivalent for LLHD. 

\begin{theorem}\label{thm:LLHD:WD2RD2}
Suppose $n\geq 2$, $\textbf{v}$ is two-dimensional, both entries of $\textbf{v}$ are coprime to $n$, $\boldsymbol{\delta} \in \mathbb{Z}^d$, and $\textbf{L}(n,\textbf{v},\boldsymbol{\delta})$ is an LLHD in \eqref{eqn:LLHD}. Then
\( c_{\text{WS}}\{\textbf{L}(n,\textbf{v},\boldsymbol{\delta})\} = c_{\text{RS}}\{\textbf{L}(n,\textbf{v},\boldsymbol{\delta})\}. \)
\end{theorem}

We propose to compute $c_{\text{WS}}\{\textbf{L}(n,\textbf{v},\boldsymbol{\delta})_{k,l}\}$ for LLHDs using the Gaussian algorithm outlined in Algorithm~\ref{alg:WS2}, which requires at most $\log_3(2n^2)+2$ iterations~\citep[Figure~2.3]{Bremner:2011}. 
Consequently, it requires a total of $O\{log(n)d^2\}$ operations to compute $c_{\text{WS2}}\{\textbf{L}(n,\textbf{v},\boldsymbol{\delta})\}$ for one LLHD. 
Furthermore, if $\textbf{v}$ and $\tilde{\textbf{v}}$ differ in only one entry, it requires $O\{\log(n)d\}$ operations to compute $c_{\text{WS2}}\{\textbf{L}(n,\tilde{\textbf{v}},\boldsymbol{\delta})\}$ from $c_{\text{WS2}}\{\textbf{L}(n,\textbf{v},\boldsymbol{\delta})\}$. 
As a result, Algorithm~\ref{alg:LLHD} takes a total of $O\{ T \log(n) d \}$ operations, which is dramatically lower than any of the preceding criteria.
Note that for large $n$, the evaluation of $c_{\text{WS2}}\{\textbf{L}(n,\textbf{v},\boldsymbol{\delta})\}$ is dramatically faster than generating the points of $\textbf{L}(n,\textbf{v},\boldsymbol{\delta})$, which requires $O(nd)$ operations. 

\begin{algorithm}[!t]
\caption{Compute the $c_{\text{WS}}(L)$ for two-dimensional designs $\textbf{L}(n,\textbf{v},\boldsymbol{\delta})$}\label{alg:WS2}
\begin{algorithmic}[1]
\STATE \textbf{Input:} sample size $n$, a two-vector $\textbf{v}$ that generates the LLHD.
\STATE Find the $\bar{\textbf{v}} \in \{1\} \times \{1,\ldots,n-1\}$ such that $\textbf{L}(n,\textbf{v},\boldsymbol{\delta}) = \textbf{L}(n,\bar{\textbf{v}},\boldsymbol{\delta})$ using the Euclidean Algorithm;
\STATE Initialize $\textbf{a} \leftarrow \bar{\textbf{v}}$ and $\textbf{b} \leftarrow (0,n)$;
\STATE Let $ \textbf{b} \leftarrow \textbf{b} - \lceil \textbf{a} \cdot \textbf{b} / \| \textbf{a} \|^2 \rfloor \textbf{a} $;
\WHILE{$\| \textbf{b} \| < \| \textbf{a} \|$} 
\STATE Let $(\textbf{a},\textbf{b}) \leftarrow (\textbf{b},\textbf{a})$; 
\STATE Let $ \textbf{b} \leftarrow \textbf{b} - \lceil \textbf{a} \cdot \textbf{b} / \| \textbf{a} \|^2 \rfloor \textbf{a} $;
\ENDWHILE
\STATE \textbf{Return} $\|\textbf{a}\|$. 
\end{algorithmic}
\vspace*{-2pt}
\end{algorithm}

Another common space-filling criterion is the fill distance (FD), defined by $c_{\text{FD}}(\textbf{X}) = \sup_{\textbf{z}\in[0,1]^d} \allowbreak \min_{i=1}^n \|\textbf{z}-\textbf{x}_i\|$, where lower values are better. 
Its wrap-around version is $c_{\text{WF}}(\textbf{X}) = \sup_{\textbf{z}\in[0,1]^d} \allowbreak \min_{i=1}^n |w(\textbf{z}-\textbf{x}_i)|$. 
\citet{Haaland:2018}, \citet{Wang:2018}, and \citet{Tuo:Wang:2020} showed that designs with both low FD and RS are appealing for Gaussian process emulation of computer experiments, and that the two criteria are only weakly contradictory to each other. 
\citet{ELIAS2020102900} recommended using the WF for the numerical integration problem. 
Despite their evident usefulness, FD and WF are not widely used in constructing space-filling designs because computing the FD or WF is extremely time-consuming for usual designs. 
Nevertheless, Theorem~\ref{thm:WF2} below shows that for LLHDs, we can compute the wrap-around bivariate fill distance (WF2), 
\( c_{\text{WF2}}(\textbf{X}) = \sum_{k < l} c_{\text{WF}}(\textbf{X}_{k,l}), \)
using $O\{log(n)d^2\}$ operations. 
Furthermore, if $\textbf{v}$ and $\tilde{\textbf{v}}$ differ in only one entry, it requires $O\{\log(n) d\}$ operations to compute $c_{\text{WF2}}\{\textbf{L}(n,\tilde{\textbf{v}},\boldsymbol{\delta})\}$ from $c_{\text{WF2}}\{\textbf{L}(n,\textbf{v},\boldsymbol{\delta})\}$. 
Consequently, for constructing designs with extremely large $n$, we recommend using WS2 or WF2. 

\begin{theorem}\label{thm:WF2}
Suppose $\textbf{v}$ is two-dimensional, both entries of $\textbf{v}$ are coprime to $n$, $\boldsymbol{\delta} \in \mathbb{Z}^d$, $\textbf{L}(n,\textbf{v},\boldsymbol{\delta})$ is an LLHD in \eqref{eqn:LLHD}, and the $\textbf{a}$ and $\textbf{b}$ are computed from the end of Algorithm~\ref{alg:WS2}. Then
\( c_{\text{WS}}\{\textbf{L}(n,\textbf{v},\boldsymbol{\delta})\} = [\{z^2+(z^2-|y|+y^2)^2\}^{1/2}/(2z)] \|\textbf{a}\|, \)
where $y = \textbf{a} \cdot \textbf{b} / \| \textbf{a} \|^2$ and $z = \|\textbf{b}- y \textbf{a}\|/\|\textbf{a}\|$. 
\end{theorem}

\subsection{Sliced LLHD} 

A sliced Latin hypercube design (SLHD) is a design that can be partitioned into several sub-LHDs~\citep{SLHD}. 
Sliced designs are useful in multi-fidelity experiments, sequential experiments, experiments with a qualitative variable, and model validation~\citep{SRSPD}. 
Suppose $s$ is a positive integer that divides $n$. 
Let
\begin{equation*}\label{eqn:SLLHD}
 \textbf{L}_j(n,\textbf{v},\boldsymbol{\delta}) = \{ \textbf{z} + i\textbf{v}/n + \boldsymbol{\delta}/n + \textbf{1}_d/(2n) : \textbf{z} \in \mathbb{Z}^p, i \in \mathbb{Z}, i - \lfloor i/s \rfloor s = j \} \cap [0,1]^d .
\end{equation*}
It is clear that $\textbf{L}_0, \dots, \textbf{L}_{s-1}$ partition $\textbf{L}$. 
Theorem~\ref{thm:SLHD} below demonstrates that all LLHDs are SLHDs, provided that $n$ is not a prime number.

\begin{theorem}\label{thm:SLHD}
Suppose $s$ divides $n$ and $\textbf{L}(n,\textbf{v},\boldsymbol{\delta})$ is an LLHD. 
Then $\textbf{L}_j(n,\textbf{v},\boldsymbol{\delta}) = \textbf{L}\{n/s,\textbf{v}, \allowbreak \boldsymbol{\delta}/s-1/2+1/(2s)+j\textbf{v}/s\}$ for any $j$ and the $\textbf{L}_j$'s have the same WS, WA, WP, WD, WS2, and WF2 values. 
\end{theorem}

Indicated by Theorem~\ref{thm:SLHD}, replacing $c\{\textbf{L}(n,\textbf{v},\boldsymbol{\delta})\}$ with a criterion that reflects space-filling properties of both $\textbf{L}(n,\textbf{v},\boldsymbol{\delta})$ and $\textbf{L}_0(n,\textbf{v},\boldsymbol{\delta})$, we can use Algorithm~\ref{alg:LLHD} to optimize sliced LLHDs. 
Optimizing sliced LLHDs should be easier than optimizing sliced OLHDs because, in the case of sliced LLHDs, we need to simultaneously optimize two designs, whereas for sliced OLHDs, we must optimize $s+1$ designs simultaneously~\citep{Ba:2015}.

\subsection{Regularly repeated designs and large-scale emulation}

Parallel to the definition of LHD, 
a design $\textbf{X}$ is said to contain an $m$-point local LHD in $\prod_{k=1}^d [l_k,u_k]$ if 
each column of $\textbf{X} \cap \prod_{k=1}^d [l_k,u_k]$ is a permutation of the points $l_k + (u_k-l_k)/(2m), l_k + 3(u_k-l_k)/(2m), \ldots, u_k - (u_k-l_k)/(2m)$. 
Consider the design given by 
\begin{equation}\label{eqn:RLHD}
 \textbf{R}(n,m,\textbf{v},\boldsymbol{\delta}) = \{ \textbf{z} m/n + i\textbf{v}/n + \boldsymbol{\delta}/n + \textbf{1}_d/(2n) : \textbf{z} \in \mathbb{Z}^p, i \in \mathbb{Z} \} \cap [0,1]^d.  
\end{equation}
Theorem~\ref{thm:RLHD} below shows that $\textbf{R}(n,m,\textbf{v},\boldsymbol{\delta})$ contains $m$-point local LHDs in any hypercube within $[0,1]^p$ whose width is $m/n$. 
We thus refer to $\textbf{R}(n,m,\textbf{v},\boldsymbol{\delta})$ as a regularly repeated lattice-based Latin hypercube design (RLHD). 

\begin{theorem}\label{thm:RLHD}
Suppose each entry of $\textbf{v}$ is coprime to $m$, $m \leq n$, $\boldsymbol{\delta} \in \mathbb{Z}^d$, and $\textbf{l} = (l_1,\ldots,l_d) \in \{0,1/n,\ldots,1-m/n\}^d$. 
Then 
\( \textbf{R}(n,m,\textbf{v},\boldsymbol{\delta}) \cap \prod_{k=1}^d [l_k,l_k+m/n] =  \textbf{L}(m,\textbf{v},\boldsymbol{\delta}-\textbf{l}n) (m/n) + \textbf{l}, \)
and $\textbf{R}(n,m,\textbf{v},\boldsymbol{\delta}) \cap \prod_{k=1}^d [l_k,l_k+m/n]$ is an $m$-point LHD in $\prod_{k=1}^d [l_k,l_k+m/n]$. 
\end{theorem}

Let $\dot{\textbf{L}}(m,\textbf{v}) = \textbf{L}(m,\textbf{v},0) (m/n) - \textbf{1}_d/(2n) = \{ \textbf{z}m/n + i\textbf{v}/n : \textbf{z} \in \mathbb{Z}^p, i \in \mathbb{Z} \}$ denote a lattice that contains $0_d$. 
Theorem~\ref{thm:tran} below shows that a fraction of local designs are translations of each other. 

\begin{theorem}\label{thm:tran}
Suppose each entry of $\textbf{v}$ is coprime to $m$, $m \leq n$, $q\leq n$, $\boldsymbol{\delta} \in \mathbb{Z}^d$, and $\textbf{l}, \tilde{\textbf{l}} \in \{0,1/n,\ldots,1-q/n\}^d$. 
Then 
\( \textbf{R}(n,m,\textbf{v},\boldsymbol{\delta}) \cap \prod_{k=1}^d [l_k,l_k+q/n] = ( \textbf{R}(n,m,\textbf{v},\boldsymbol{\delta}) \cap \prod_{k=1}^d [\tilde l_k, \allowbreak \tilde l_k+q/n] ) + \textbf{l} - \tilde{\textbf{l}} \)
if and only if $\textbf{l} - \tilde{\textbf{l}} \in \dot{\textbf{L}}(m,\textbf{v}) $. 
\end{theorem}

\begin{figure}[!t]
\centering
\includegraphics[width=0.6\textwidth]{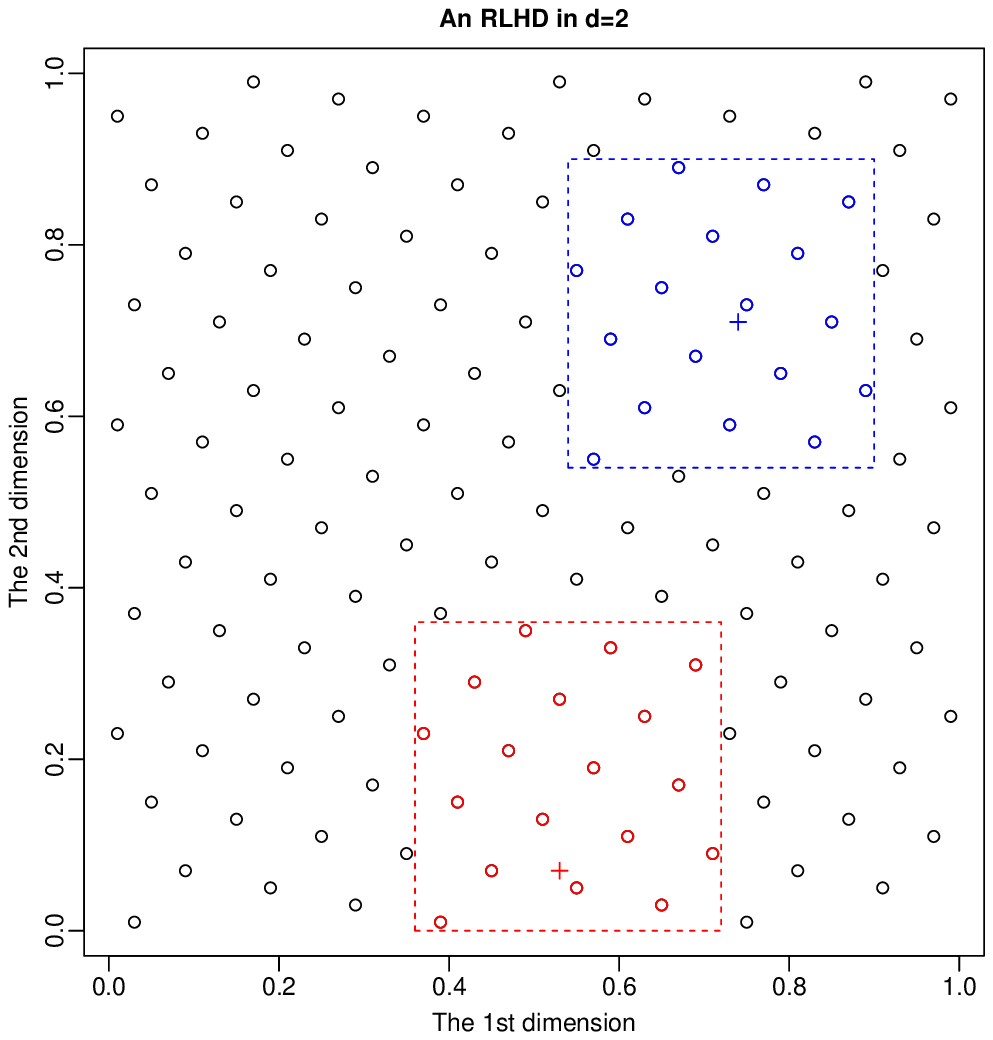}
\caption{An RLHD in two dimensions, showing the designs points (circles) and the two local LHDs (boxs with dashed lines) for two positions to make prediction (pluses). \label{fig:RLHD:show}}
\end{figure}

Although Theorem~\ref{thm:tran} applys to any $q \leq n$, in practice we usually set $q=m$ because otherwise the local designs are not LHD. 
By exploiting Theorems~\ref{thm:RLHD} and~\ref{thm:tran}, we propose using Algorithm~\ref{alg:GP:RLHD} to apply RLHDs to the large-scale Gaussian process emulation problem. 
In Step~\ref{step:localize}, we find the optimal $\dot{\textbf{l}}_i$ such that the point $\dot{\textbf{x}}_i$ is positioned as close as possible to the center of the hypercube $\prod_{k=1}^d [\dot l_k, \dot l_k + m/n]$. 
Since $\dot{\textbf{l}}_i \in \dot{\textbf{L}}(m,\textbf{v})$ for all $i$, the correlation matrices for all local GPMs are identical. 
Therefore, we only need to compute their inverse once. 
For illustration, Figure~\ref{fig:RLHD:show} displays a two-dimensional RLHD, $R\{50,18,(1,7),(13,12)\}$, which is obtained by minimizing the WS using Algorithm~\ref{alg:LLHD}, along with two local LHDs contained within it, each derived from a specific input position where we intend to make predictions. 
We observe that both local LHDs are centered around their respective input positions, and that the two LHDs are translations of one another.

\begin{algorithm}[!t]
\caption{Fit Gaussian process models and make predictions using an RLHD}\label{alg:GP:RLHD}
\begin{algorithmic}[1]
    \STATE \textbf{Input:} An RLHD $\textbf{R}(n,m,\textbf{v},\boldsymbol{\delta})$, outputs corresponding to the $\textbf{R}(n,m,\textbf{v},\boldsymbol{\delta})$, and a list of input values $\dot{\textbf{x}}_1,\ldots,\dot{\textbf{x}}_{\dot n}$ to make predictions.
\STATE Compute the inverse of the corvariance matrix corresponding to the $m$-point local design $\textbf{L}(m,\textbf{v},\boldsymbol{\delta}) (m/n)$; \label{step:invert} 
\STATE Generate the points of $\dot{\textbf{L}}(m,\textbf{v}) \cap [0,1-m/n]^d$;
\FOR{$i$ from 1 to $\dot n$} 
\STATE Find the $\dot{\textbf{l}}_i \in \dot{\textbf{L}}(m,\textbf{v}) (m/n) \cap [0,1-m/n]^d$ that minimizes $\| \dot{\textbf{x}}_i - \{\dot{\textbf{l}}_i + m/(2n)\} \|$; \label{step:localize}
\STATE Predict $f(\dot{\textbf{x}}_i)$ using outputs corresponding to $\textbf{L}(m,\textbf{v},\boldsymbol{\delta}-\dot{\textbf{l}}_i n) (m/n) + \dot{\textbf{l}}_i$. \label{step:predict}
\ENDFOR
\end{algorithmic}
\vspace*{-2pt}
\end{algorithm}

Since emulation accuracy is closely related to space-filling properties of local designs, the local LLHD should be optimized. 
We recommend using Algorithm~\ref{alg:LLHD} to find the optimal $\textbf{v}$ by setting $n \leftarrow m$ and $d \leftarrow d$ as inputs for the algorithm. 
Based on our experience, the size of $\textbf{R}(n,m,\textbf{v},\boldsymbol{\delta})$ is usually very close to $n^d m^{-d+1}$, i.e., the averaged size indicated by Theorem~\ref{thm:RLHD:n} below. 
Thus, we can determine the $n$ of RLHD based on the size of local designs and the target number of design points.

\begin{theorem}\label{thm:RLHD:n}
For any $n>m>1$ and $\textbf{v}$ whose entries are coprime to $m$, 
for $\boldsymbol{\delta}$ randomly generated from the uniform distribution in $\{0,\ldots,m-1\}^d$, 
the expected number of design points for $\textbf{R}(n,m,\textbf{v},\boldsymbol{\delta})$ is $n^d m^{-d+1}$. 
\end{theorem}

\section{Numerical comparison} 
\label{sec:comp}

\begin{figure}[t!]
\centering
\includegraphics[width = 0.49\textwidth]{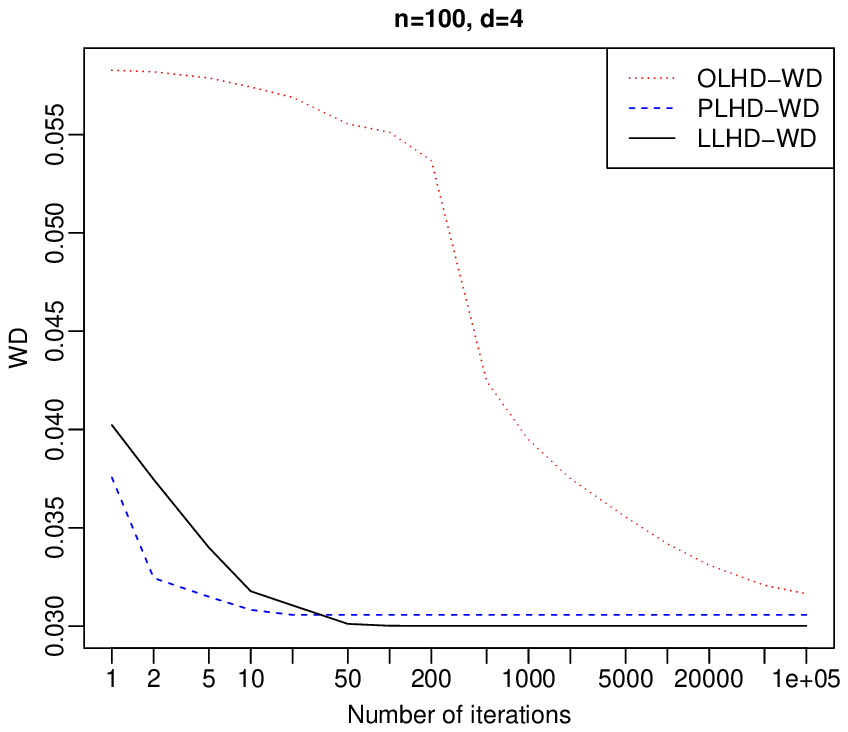}
\includegraphics[width = 0.49\textwidth]{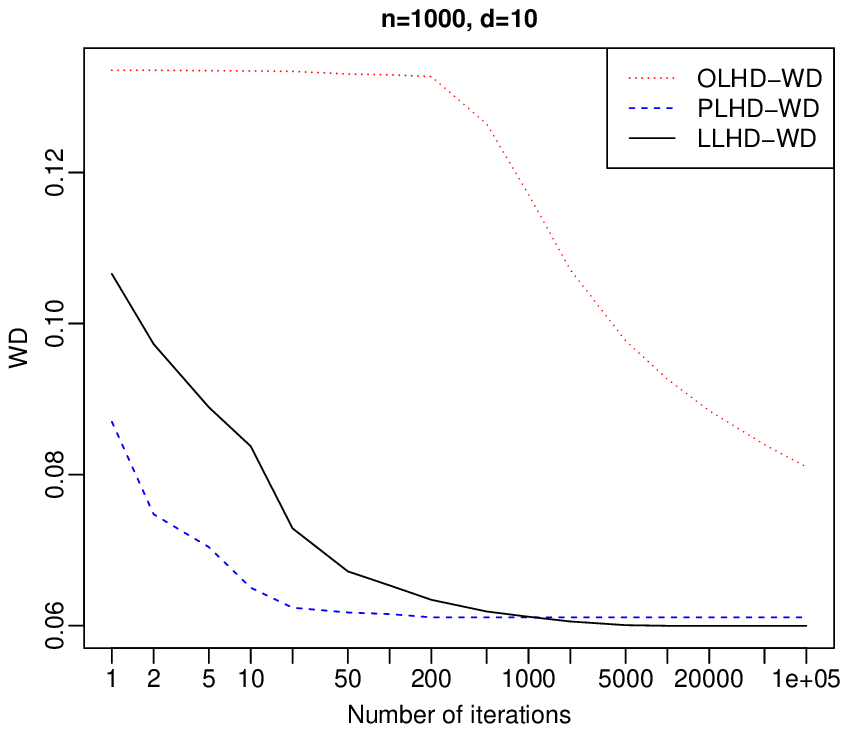}

\includegraphics[width = 0.49\textwidth]{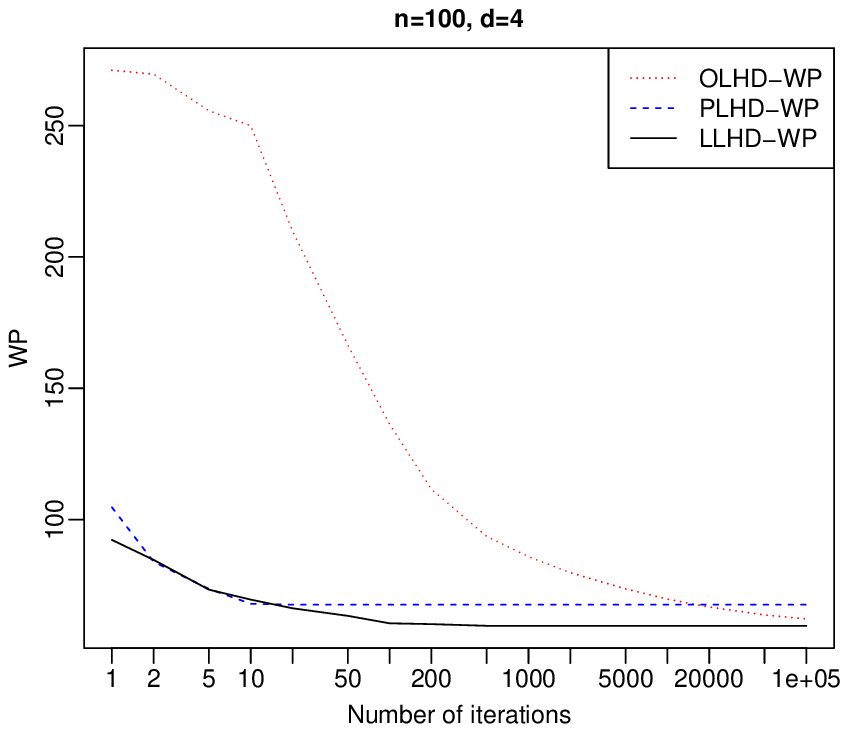}
\includegraphics[width = 0.49\textwidth]{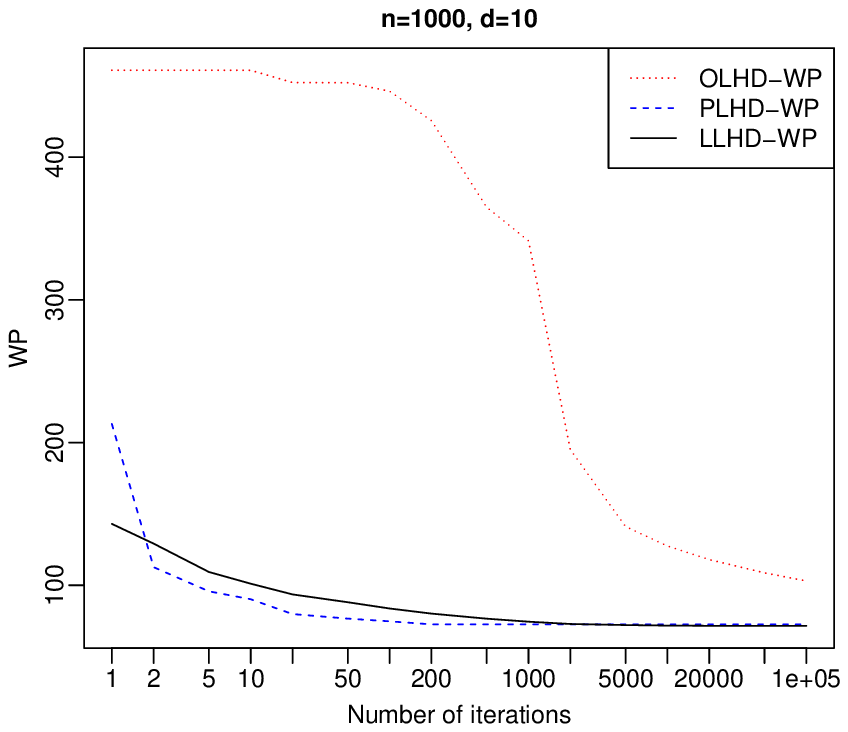}

\includegraphics[width = 0.49\textwidth]{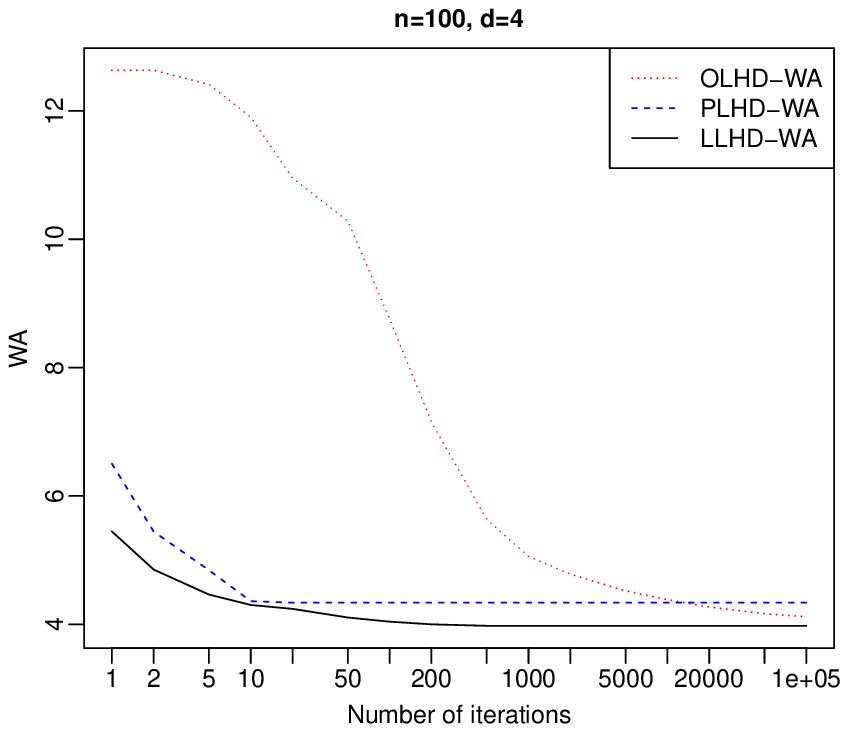}
\includegraphics[width = 0.49\textwidth]{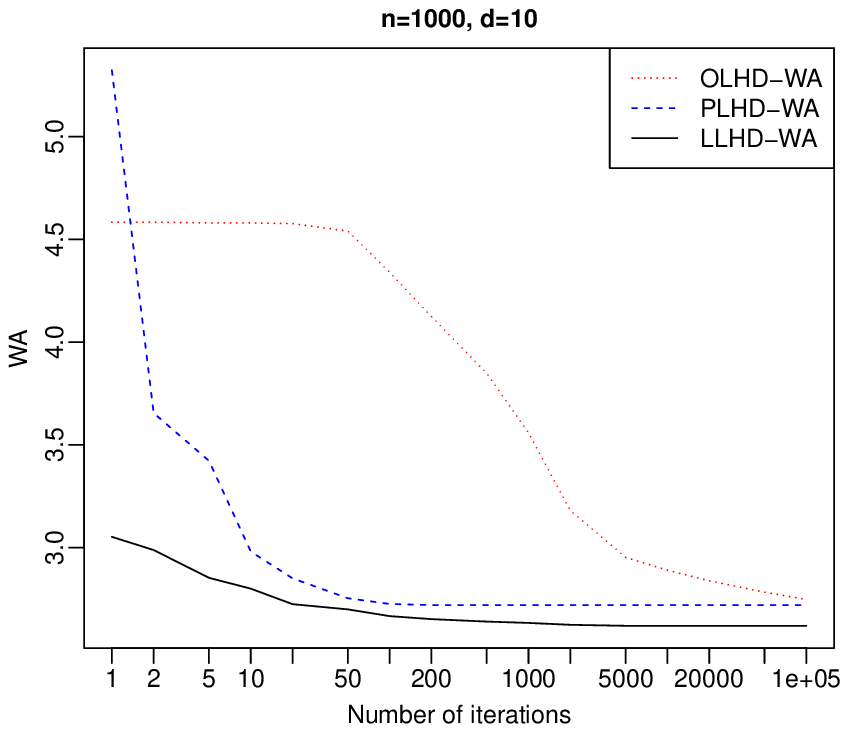}
\caption{The WD (top), WP (middle), and WA (bottom) values versus number of iterations $T$ for designs in $n=100$, $d=4$ (left) and $n=1000$, $d=10$ (right). \label{fig:conv}}
\end{figure}

In this section, we present numerical results comparing different types of designs. 
First, we examine the number of iterations required for the optimization schemes to converge. 
We consider two cases: $(n=100, d=4)$ and $(n=1000, d=10)$. 
In the top panels of Figure~\ref{fig:conv}, we plot the WD values against the number of iterations from optimizing the WD criterion for LHD, PLHD, and LLHD. 
Similarly, the middle and bottom panels display the WP and WA values, respectively, for the algorithms optimizing these criteria. 
All results are averaged over 10 random starts. 

For $n=100$ and $d=4$, approximately 500 iterations are required to find the optimal LLHD. 
For $n=1000$ and $d=10$, about 5000 iterations are needed. 
From additional numerical results not shown here, it appears that $5p(n)d$ iterations are generally sufficient for the convergence of LLHD. 
In contrast, for all cases, the algorithm for OLHD does not converge to the optimal LHD within $10^5$ iterations. 
This suggests that when $n\geq 100$, the search space for unrestricted LHDs is too large to reach the global or even a local optimum. 
From the fact that LLHDs converge much more quickly than OLHDs, we conclude that it takes less computation to find the optimal LLHD. 
Indicated from the fact that LLHDs have much lower criterion values than OLHDs at $T=1$, random LLHDs are more space-filling than random LHDs. 
Persumably because power generators are on average better than random generators for the WD criterion~\citep{Korobov:1959}, PLHDs at $T=1$ outperform LLHDs at $T=1$ for the WD criterion, 
However, LLHDs at $T=1$ outperform PLHDs on WP and WA, showing power generators have no advantage on other criteria. 
Although the optimal PLHD can be obtained more quickly, it is inferior to the optimal LLHD, suggesting that the optimal LLHD is not a PLHD. 
We thus conclude that restricting the search to PLHDs is overly limiting. 
Because LLHDs consistently outperform OLHDs and PLHDs after $10^5$ iterations, we conclude that it is reasonable to limit the search to LLHDs.

\begin{figure}[t!]
\centering
\includegraphics[width = 0.49\textwidth]{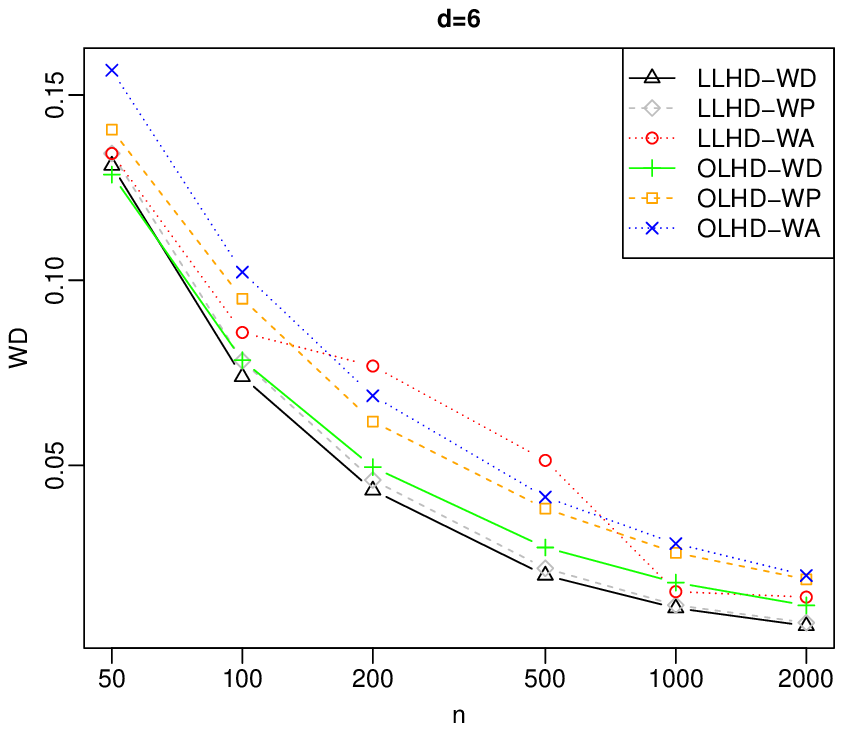}
\includegraphics[width = 0.49\textwidth]{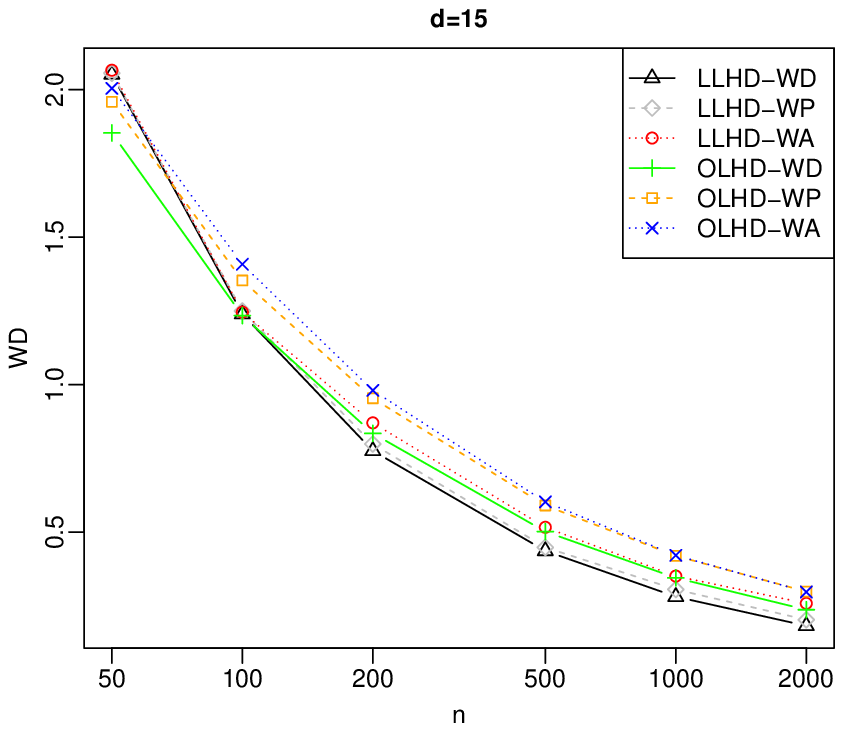}

\includegraphics[width = 0.49\textwidth]{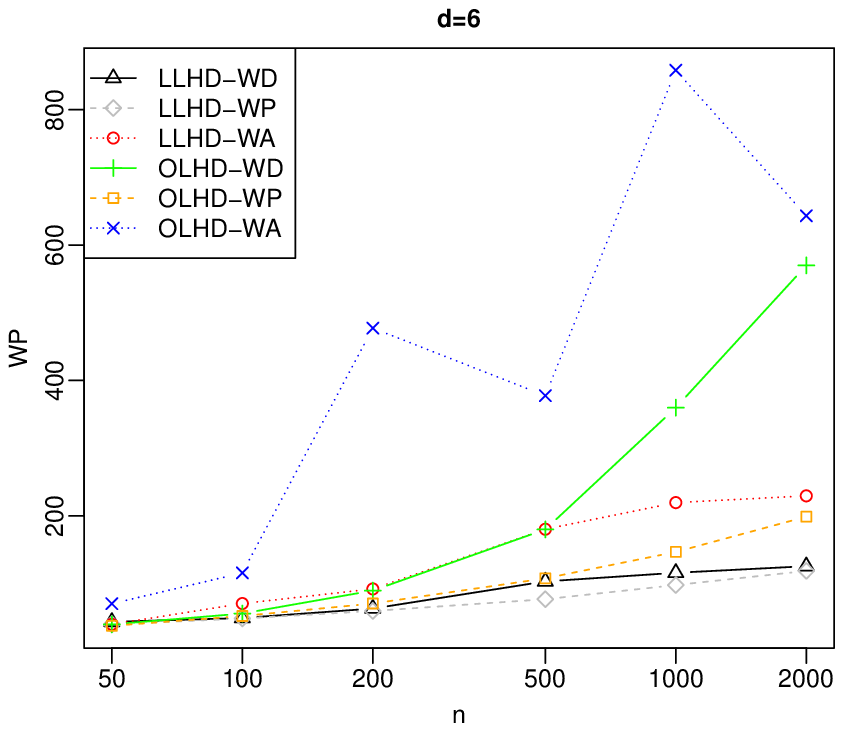}
\includegraphics[width = 0.49\textwidth]{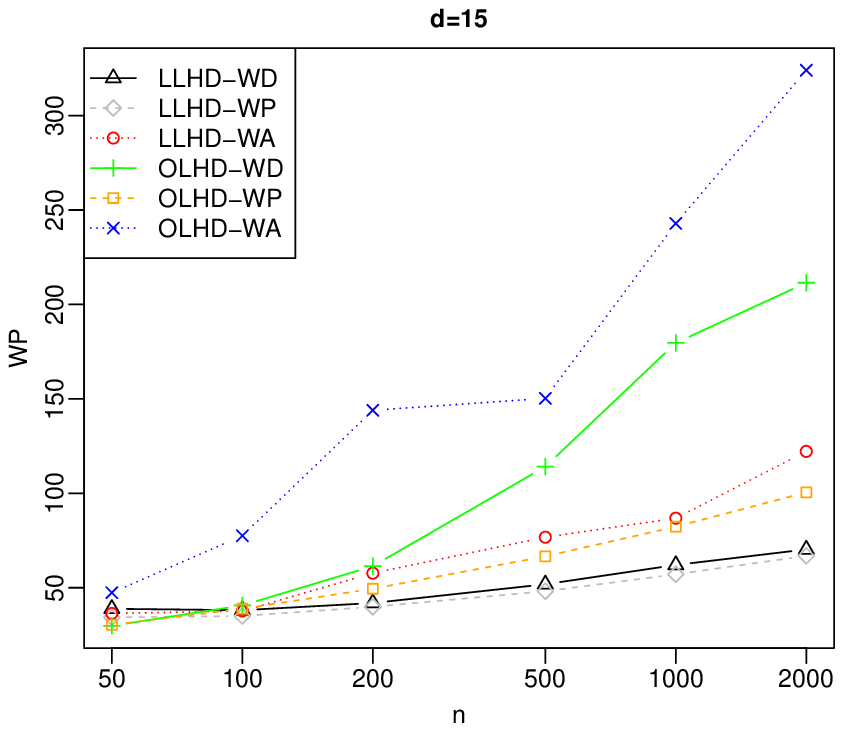}

\includegraphics[width = 0.49\textwidth]{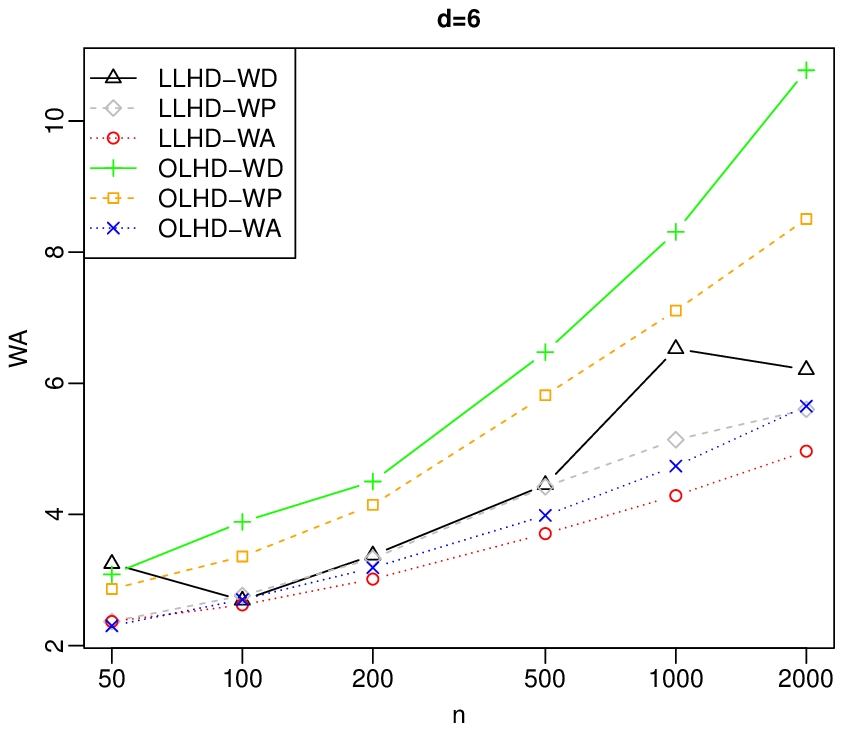}
\includegraphics[width = 0.49\textwidth]{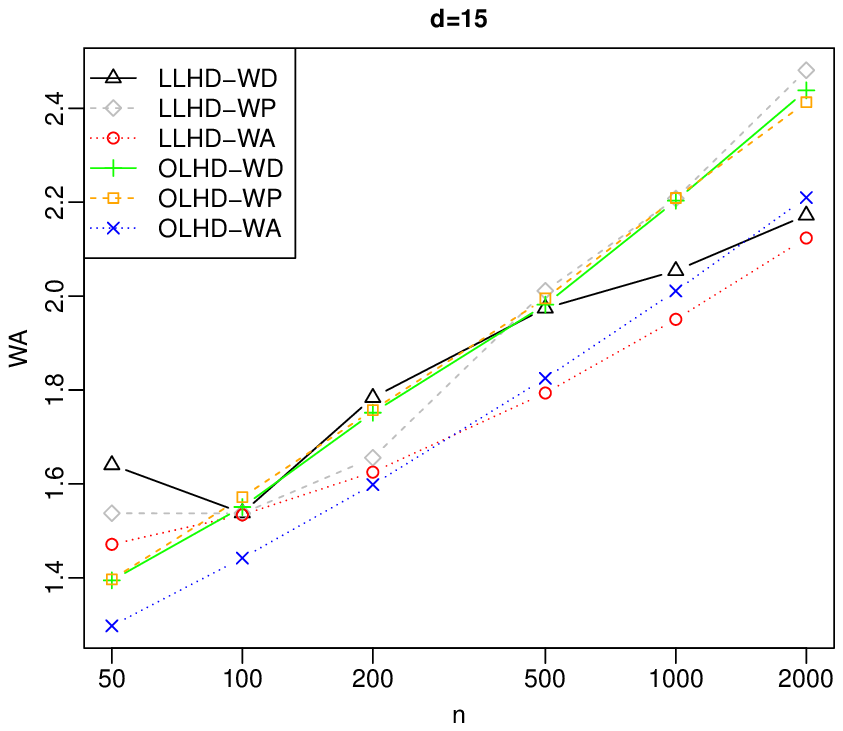}
\caption{The WD (top), WP (middle), and WA (bottom) values versus $n$ for LLHDs and OLHDs generated by optimizing WD, WP, and WA in $d=6$ (left) and $d=15$ (right). \label{fig:prop}}
\end{figure}

Next, we compare OLHDs and LLHDs with respect to their space-filling properties after $10^5$ iterations. 
Figure~\ref{fig:prop} shows the WD, WP, and WA values for OLHD and LLHD, each optimized using different criteria. 
We draw two major conclusions from the results. 
Firstly, among the optimized LLHDs and OLHDs of the same size and optimization criterion, LLHDs are superior in the majority of cases when $n \geq 100$. 
Secondly, the optimized LLHDs excel not only on the criterion used in the optimization but also perform reasonably well on the other two criteria. 
In contrast, OLHDs tend to perform poorly on the other two criteria. 
For example, LLHDs optimized for WD (LLHD-WD) consistently outperform OLHDs optimized for WD (OLHD-WD) across all three criteria—WD, WP, and WA.
Moreover, the performance gaps on WP and WA are much larger than on WD. 
To understand why LLHDs perform well across all three criteria, we generate 1000 random LLHDs for $n=100$ and $d=4$ and compute the sample correlation between the WD, WP, WS, and WS2 values. 
Surprisingly, we find that the four criteria are all positively correlated. 
In fact, all pairwise correlations, except between WS and WS2, are at least 0.37. 
See Table~\ref{tab:corr:100:4} for the details. 
This suggests that by optimizing just one criterion, we may be able to enhance all space-filling properties of LLHDs. 
Conversely, random LHDs do not exhibit this phenomenon, as the second-highest correlation is only 0.13. 
This explains why OLHDs perform poorly on criteria other than the one used in the optimization algorithm. 
We repeat the comparison for $n=1000$ and $d=10$, and the results are similar, as shown in Table~\ref{tab:corr:1000:10}. 
Clearly, LLHDs exhibit substantially better overall space-filling properties than OLHDs for $n \geq 100$.

\begin{table}[!t]
\centering
\caption{Correlations between criteria for random LHDs and random LLHDs in $n=100$ and $d=4$}
\vspace{.3cm}
\begin{tabular}{|c|cccc|}
\hline
LHD&  WS  &  WP  &  WD  &  WS2 \\
\hline
WS  & 1    & 0.77 & 0.08 & 0.07 \\
WP  & 0.77 & 1    & 0.08 & 0.13 \\
WD  & 0.08 & 0.08 & 1    & 0.02 \\
WS2 & 0.07 & 0.13 & 0.02 & 1    \\
\hline
\end{tabular} 
\quad\quad
\begin{tabular}{|c|cccc|}
\hline
LLHD&  WS  &  WP  &  WD  &  WS2 \\
\hline
WS  & 1    & 0.37 & 0.58 & 0.06 \\
WP  & 0.37 & 1    & 0.67 & 0.49 \\
WD  & 0.58 & 0.67 & 1    & 0.88 \\
WS2 & 0.06 & 0.49 & 0.88 & 1    \\
\hline
\end{tabular}
\label{tab:corr:100:4}
\end{table}

\begin{table}[!t]
\centering
\caption{Correlations between criteria for random LHDs and random LLHDs in $n=1000$ and $d=10$}
\vspace{.3cm}
\begin{tabular}{|c|cccc|} 
\hline
LHD&  WS  &  WP  &  WD  &  WS2 \\
\hline
WS  & 1    & 0.07 & 0.08 &-0.04 \\
WP  & 0.07 & 1    & 0.01 &-0.03 \\
WD  & 0.08 & 0.01 & 1    & 0.01 \\
WS2 &-0.04 &-0.03 & 0.01 & 1    \\
\hline
\end{tabular}
\quad\quad
\begin{tabular}{|c|cccc|}
\hline
LLHD&  WS  &  WP  &  WD  &  WS2 \\
\hline
WS  & 1    & 0.22 & 0.06 &-0.01 \\
WP  & 0.22 & 1    & 0.50 & 0.40 \\
WD  & 0.06 & 0.50 & 1    & 0.77 \\
WS2 &-0.01 & 0.40 & 0.77 & 1    \\
\hline
\end{tabular}
\label{tab:corr:1000:10}
\end{table}


\begin{figure}[t!]
\includegraphics[width = 0.49\textwidth]{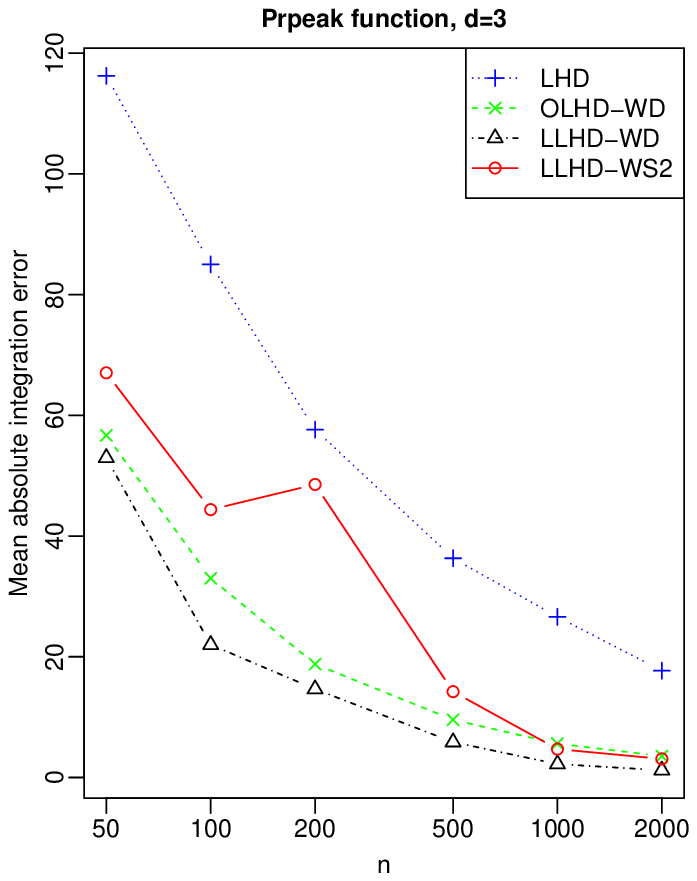}
\includegraphics[width = 0.49\textwidth]{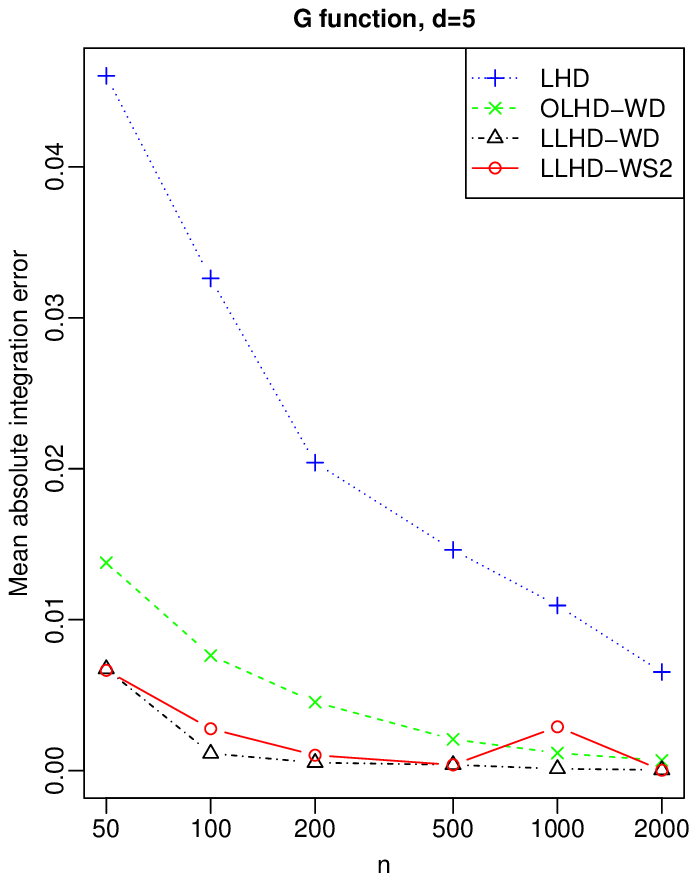}

\includegraphics[width = 0.49\textwidth]{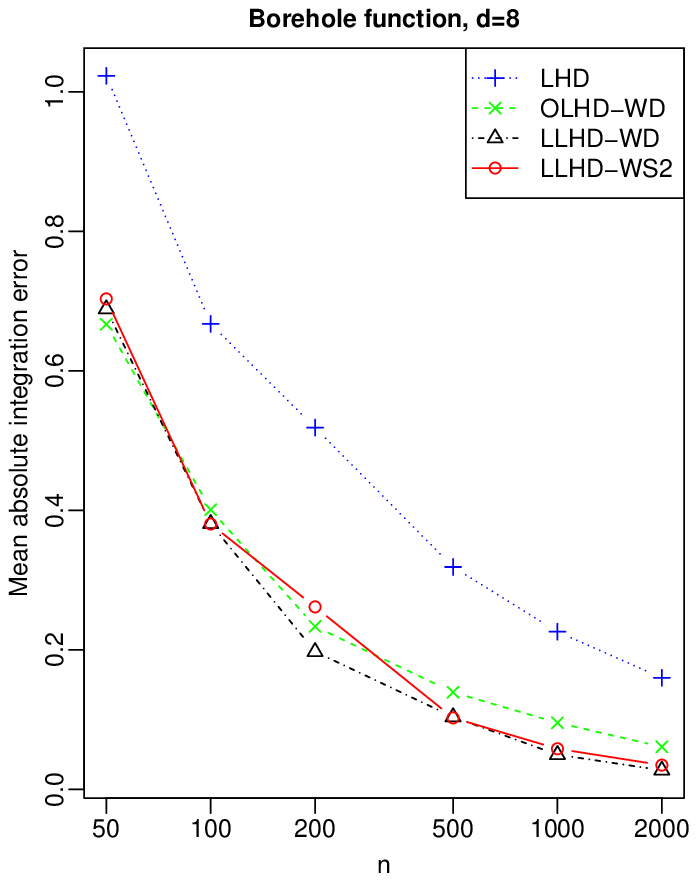}
\includegraphics[width = 0.49\textwidth]{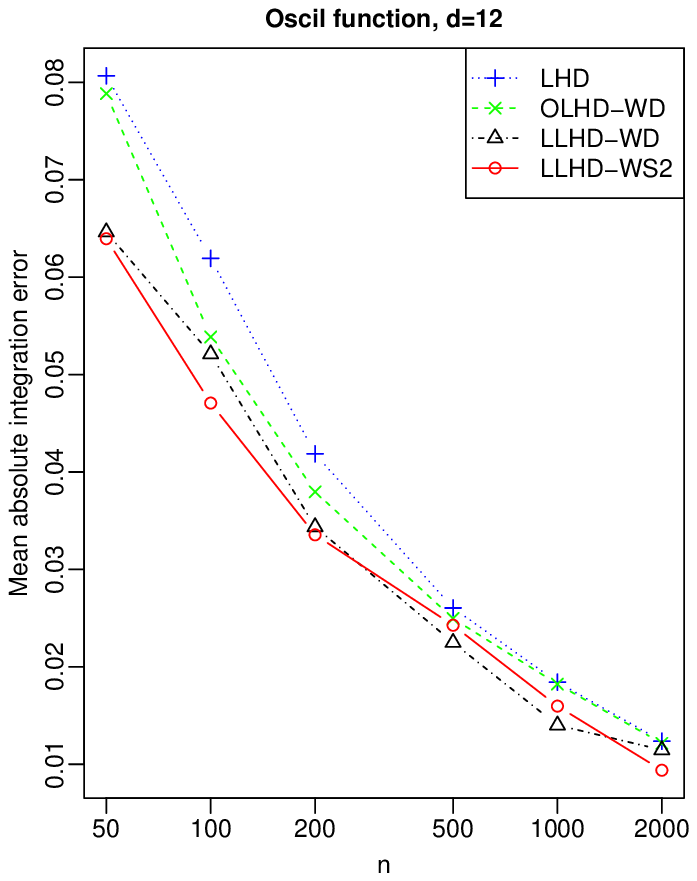}
\caption{Mean absolute integration errors versus $n$ for the 3-dimensional Prpeak function (topleft), the 5-dimensional G-function (topright), the 8-dimensional Borehole function (bottomleft), and the 12-dimensional Oscil function (bottomright). \label{fig:inte}}
\end{figure}

To evaluate the performance of different designs in uncertainty quantification problems, we compare their effectiveness in estimating the mean output value of computer simulations. 
Figure~\ref{fig:inte} illustrates the mean absolute integration error using random LHD (LHD), LHD with optimized WD (OLHD-WD), LLHD with optimized WD (LLHD-WD), and LLHD with optimized WS2 (LLHD-WS2) for four test functions: the 3-dimensional Prpeak function, the 5-dimensional G-function, the 8-dimensional Borehole function, and the 12-dimensional Oscil function~\citep{Bingham:library}. 
The results clearly indicate that LLHD-WD performs the best, while LLHD-WS2 is only marginally worse than LLHD-WD in more than half of the cases. 
This demonstrates that the superior space-filling properties of LLHDs translate into better uncertainty quantification accuracy. 
Moreover, when $n$ becomes very large, LLHD-WS2 can be used as it offers reasonable performance while requiring significantly less computational effort. 

Next, we compare design methods in training PINNs. 
We consider two PDEs~\citep{wu2023comprehensive}. 
The first is the Burgers' equation given by 
\[ \left\{ \begin{array}{ll}
\partial u(x,t) / \partial t + u(x,t) \partial u(x,t) / \partial x = (0.01/\pi) \partial^2 u(x,t) / \partial x^2, &  x \in [-1, 1], t \in [0, 1],\\
 u(x, 0) = -\sin(\pi x), & x \in [-1, 1], \\
 u(-1, t) = u(1, t) = 0, &  t \in [0, 1], 
\end{array} \right. \]
where $u(x,t)$ denotes the flow velocity at position $x$ and time $t$. 
The true solution of the Burgers' equation has a sharp front at $x = 0$~\citep{wu2023comprehensive}. 
The second is the wave equation given by 
\[ \left\{\begin{array}{ll}
    \partial^2 u(x,t) / \partial t^2 - 4 \partial^2 u(x,t) / \partial x^2 = 0, &  x \in [0, 1], t \in [0, 1], \\
    u(0, t) = u(1,t) = 0, &  t \in [0, 1], \\
    u(x, 0) = \sin(\pi x) + \sin(4 \pi x)/2, &  x \in [0, 1], \\
    \partial u(x, 0) / \partial t = 0, &  x \in [0, 1],
\end{array}\right. \]
where $u(x,t)$ denotes the displacement of the wave at position $x$ and time $t$.
The exact solution of the wave equation has an analytical solution, 
$u(x,t) = \sin(\pi x) \cos(2 \pi t) + 0.5 \sin(4 \pi x) \cos(8 \pi t)$, showing its multi-scale behavior in both spatial and time dimensions.

\begin{figure}[t!]
\centering
\includegraphics[width = 0.49\textwidth]{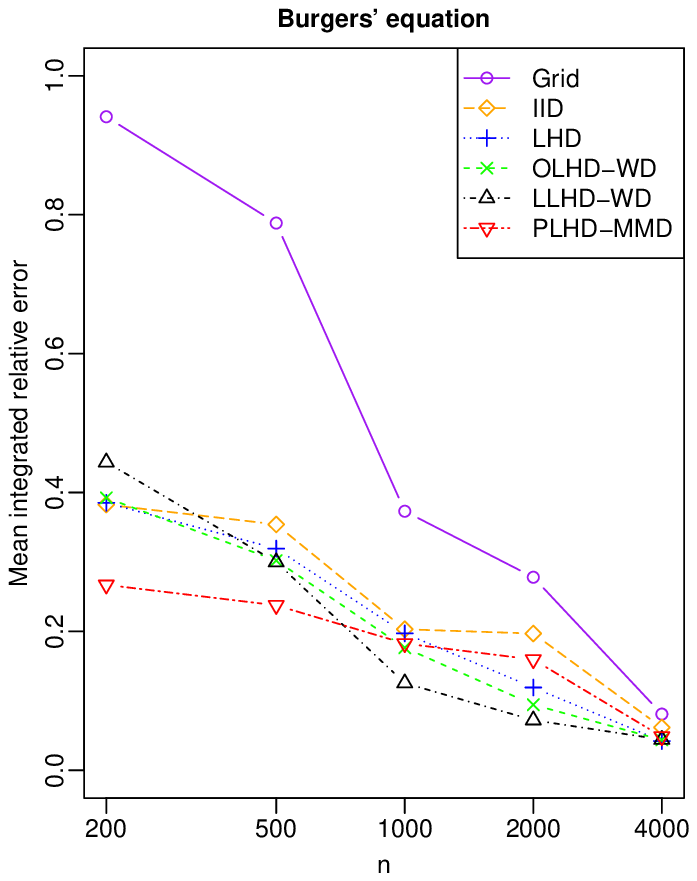}
\includegraphics[width = 0.49\textwidth]{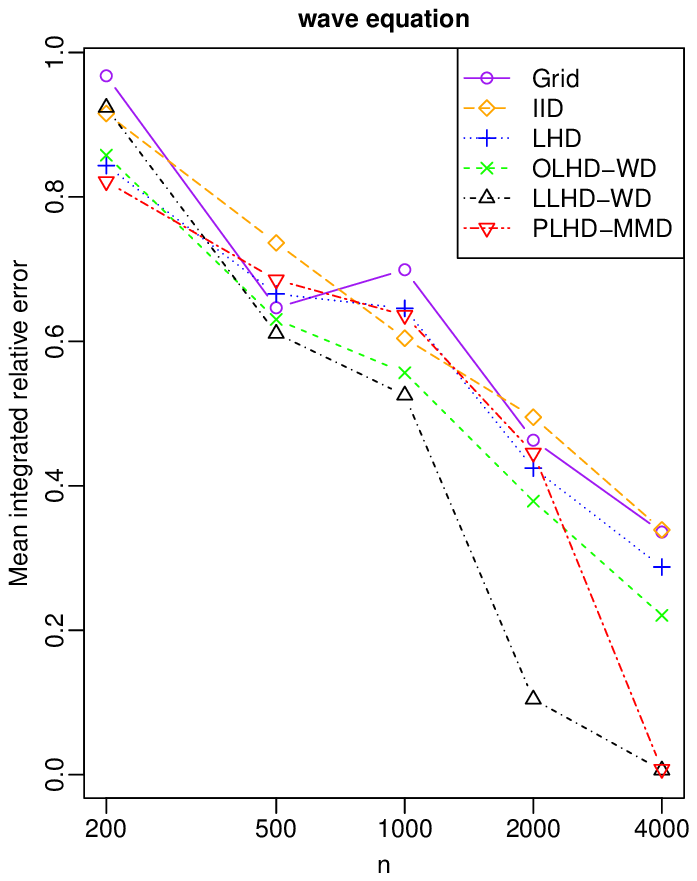}
\caption{Mean integrated relative error versus $n$ for the Burgers' equation (left) and the wave equation (right). \label{fig:pinn}}
\end{figure}

We use the method in \citet{matsubara2023good} to fit PINNs. 
The two PDEs are solved using PINN where the design points are generated using LHD, OLHD-WD, LLHD-WD, the good lattice design proposed in \citet{matsubara2023good} that exhaustively searches for the power generator that minimizes the maximum discretization error (PLHD-MMD), the full grid design (Grid), and independently and identically generated points (IID).  
Fig.~\ref{fig:pinn} presents the $L^2$ integrated relative error of the solutions. 
As can be clearly seen, LLHD-WD outperforms other methods in both PDEs when $n \geq 10^3$. 
This is presumably because the loss function produced using the LLHD-WD method has excellent numerical integration accuracy. 

Finally, we compare the design and modeling strategies for large-scale emulations. 
We select three test functions that are challenging to fit due to sharp changes in the simulation outputs: the three-dimensional Ackley function, the four-dimensional Shekel function, and the six-dimensional Emichalewicz function~\citep{Mullen:2014}. 
We compare the performance of our proposed RLHD approach against four other methods: PD, NN, laGP, and SG, which we will explain below. 
For all approaches except SG, we assume that a total of $n=4000$, $8000$, and $32000$ simulation outputs are collected for the three test functions, respectively, and local models are fitted using $m=500$ subsamples to make predictions. 
In the partitioned design approach (PD), the input space is divided into $2^d$ sub-hypercubes by partitioning each dimension into two halves. 
An LHD with $m=n/2^d=500$ points is used as the design, and a local GPM is fitted for each subregion. The prediction for a test point is made using the local model corresponding to the subregion containing the point. 
This approach requires $O(m^3 2^d) = O(n m^2)$ operations, which is smaller than $O(n^3)$ but larger than $O(m^3)$ that is required for RLHD. 
In the nearest neighbor approach (NN), the $n$ outputs are generated from a single $n$-point LHD. 
For each prediction, a local GPM is fitted using $m=500$ training samples that are closest to the input location. 
In the local approximate GPM approach (laGP)~\citep{laGP}, the $m$ nearest neighbor samples are replaced with $m$ actively learned samples. 
For both NN and laGP approaches, we use the R package laGP~\citep{R:laGP}. Both approaches require $O(m^3 n_{\text{p}})$ operations, where $n_{\text{p}}$ represents the number of predictions. 
In practice, we expect that $n_{\text{p}} > n$; otherwise, we could directly run simulations at the desired positions. Consequently, both NN and laGP approaches require at least $O(n m^3)$ operations, which are significantly more computationally expensive than PD and RLHD. 
Finally, the sparse grid approach (SG) proposed by \citet{Plumlee:2014} uses a single global model for predictions, based on all $n$ observations. 
Since SG requires the design to be a sparse grid and imposes constraints on $n$, we use $n=4089$, $8361$, and $40081$ for the three test functions, respectively, as these are the nearest allowable sizes.

We assume that the scale parameters are unknown and need to be estimated before fitting the model. 
For PD, NN, laGP, and SG, we estimate the scale parameters by maximizing the likelihood function, as done in both \citet{laGP} and \citet{Plumlee:2014}. 
Although this yields accurate scale parameters, it requires many iterations of numerical optimization, with each iteration involving inverting the correlation matrix. As a result, estimating the scale parameters demands significantly more computational effort than model fitting.
For large-scale emulation problems, we propose using smaller subdesigns to estimate the scale parameters. 
From Theorem~\ref{thm:tran}, for any integer $q$ smaller than $n$, the $\textbf{R}(n,m,\textbf{v},\boldsymbol{\delta})$ contains many $q$-point local designs that are translations of each other. 
In the RLHD approach, we randomly generate 10 local designs with $q=50$ points. We then estimate the scale parameters by maximizing the composite likelihood across the 10 designs. 
Since all 10 designs are translations of each other, only one inversion of the $q \times q$ correlation matrix is required per iteration. 
Since $O(q^3)$ is much smaller than $O(m^3)$, estimating the scale parameters in the RLHD approach requires significantly less computational effort than model fitting. 
As a result, RLHD is much more computationally efficient than PD, NN, and laGP. 

\begin{table}[!t]
\centering
\caption{RMSE for emulating three testing functions}\label{tab:emulation}
\vspace{.3cm}
\begin{tabular}{cc|ccccc}
\hline
Function & $d$ & RLHD & PD & NN & laGP & SG \\
\hline
Ackley & 3  & 9.54e-6 & 4.12e-4 & 1.31e-4 & 9.60e-5 & 2.81e-4\\
Shekel & 4  & 0.017 & 0.040 & 0.034 & 0.110 & 0.031\\
Emichalewicz & 6 & 0.478 & 0.629 & 0.510 & 0.542 & 0.611 \\
\hline
\end{tabular}

\end{table}

Table~\ref{tab:emulation} presents the root mean squared prediction error (RMSE) for predicting $1000$ randomly generated test points for each method with each test function. 
The results show that RLHD achieves the smallest RMSE. 
This is likely due to the fact that the local designs in RLHD are optimized LLHDs, whereas the local designs for PD, NN, and laGP, as well as the full design in SG, are not as space-filling. 
In conclusion, the RLHD approach is attractive both in terms of prediction accuracy and computational efficiency.

\section{Final remarks}
\label{sec:conc}

In this paper, we propose new methods for efficiently generating space-filling designs with arbitrary $n$ and $d$. 
The key innovation is the use of new shortcut formulas that reduce the computational complexity from $O(n^2)$ to $O(n)$ and $O\{\log(n)\}$. 
Because the constructed designs are appealling for numerical integration problems, they are excellent in fitting PINNs. 
Future work is needed to see if these designs are useful for other deep learning problems. 

We also introduce a method for constructing regularly repeated designs that contain numerous local space-filling LHDs as subdesigns.
These designs, when combined with a moving window type of local modeling technique, facilitate the rapid fitting of Gaussian process emulators.
Although we focus primarily on moderately time-consuming computer simulations, where hundreds or thousands of simulation trials can be afforded, our method is capable of constructing extremely large designs and can be applied to general black-box functions in other statistical fields.
The regularly repeated structure of our proposed designs stems from their lattice-based structure.
It may be possible to construct regularly repeated designs using other types of lattices that have been previously used for constructing space-filling designs, such as thinnest covering lattice~\citep{RSPD}, densest packing lattice~\citep{DPMPD}, and interleaved lattice~\citep{ILmMD,Maximin}.
This is an exciting direction for future research.



\bigskip
\begin{center}
{\large\bf SUPPLEMENTARY MATERIAL}
\end{center}

\begin{description}


\item[Proofs:] Proof of Theorems. (.pdf)
\item[R-package:] Source codes of an R package for constructing our proposed designs. (.tar.gz) 
\item[Code:] Codes to produce the numerical results. (.zip) 

\end{description}


\bibliographystyle{apalike}
\bibliography{LOLH}

\end{document}



\if1\blind
{
  \title{\bf Supplementary material for ```Optimized and regularly repeated lattice-based Latin hypercube designs for large-scale computer experiments'''}
  \author{Xu He\\
    State Key Laboratory of Mathematical Sciences (SKLMS), \\Academy of Mathematics and Systems Science, \\Chinese Academy of Sciences, hexu@amss.ac.cn, \\
Junpeng Gong\\
    School of Mathematical Sciences, University of Chinese Academy of \\Sciences and State Key Laboratory of Mathematical Sciences (SKLMS), \\Academy of Mathematics and Systems Science, \\Chinese Academy of Sciences,  gongjunpeng21@mails.ucas.ac.cn\\
and\\
Zhaohui Li\\
    State Key Laboratory of Mathematical Sciences (SKLMS), \\Academy of Mathematics and Systems Science, \\Chinese Academy of Sciences, lizh@amss.ac.cn
}
  \maketitle
} 
\fi

\if0\blind
{
  \bigskip
  \bigskip
  \bigskip
  \begin{center}
    {\LARGE\bf Supplementary material for ```Optimized and regularly repeated lattice-based Latin hypercube designs for large-scale computer experiments'''}
\end{center}
  \medskip
} \fi

\bigskip

\abstract{
In this supplementary material, we provide proofs for the corresponding theorems. } 


\begin{proof}[Proof of Theorem~1:]
We have 
\[ c_{\text{WS}}(X) 
 = \max_{i \neq j} \left[ \left\{ \sum_{k=1}^d w( (j-i)v_k/n )^2 \right\}^{-1/2} \right] 
 = \min_{i=1}^{n-1} \left[ \left\{ \sum_{k=1}^d w( iv_k/n )^2 \right\}^{-1/2} \right] ,\] 
\[ c_{\text{WA}}(X) 
 = \left[ \sum_{i < j} \left\{ \sum_{k=1}^d w( (j-i)v_k/n )^2 \right\}^{-50/2} \right]^{1/50} 
 = \left[ \frac{n}{2} \sum_{i=1}^{n-1} \left\{ \sum_{k=1}^d w( iv_k/n )^2 \right\}^{-50/2} \right]^{1/50} ,\]
\[ \
    c_{\text{WP}}(X) 
= \left[ \sum_{i < j} \left\{ \prod_{k=1}^d w( (j-i)v_k/n )^{-2} \right\} /\{n(n-1)/2\} \right]^{1/d} 
= \left[ \frac{n}{2} \sum_{i=1}^{n-1} \left\{ \prod_{k=1}^d w( iv_k/n )^{-2} \right\} \right]^{1/d},
\]

\begin{align*} 
c_{\text{WD}}(X)^2 
 &= \sum_{i,j=1}^n \prod_{k=1}^d \left\{ 1.5 - r( (j-i)v_k/n ) ) + r( (j-i)v_k/n )^2 \right\} /n^2 - (4/3)^d \\
 &= \sum_{i=1}^n \prod_{k=1}^d \left\{ 1.5 - r( iv_k/n ) +r( iv_k/n )^2 \right\} /n - (4/3)^d \\ 
 &= \sum_{i=1}^n \prod_{k=1}^d \left\{ 1.25 + w( iv_k/n -1/2 )^2 \right\} /n - (4/3)^d. 
\end{align*}
\end{proof}

\begin{proof}[Proof of Theorem~2:]
When $n=2$, $L(n,v,\delta)$ is either $\{ (1/4,1/4), (3/4,3/4) \}$ or $\{ (1/4,3/4), \allowbreak (3/4,1/4) \}$. 
In both cases, $c_{\text{RS}}\{L(n,v,\delta)\}^{-1} = c_{\text{WS}}\{L(n,v,\delta)\}^{-1} = 2^{-1/2}$. 

When $n>2$, from Lemma~1, there exists an $a = (a_1,a_2) \in \mathbb{Z}^2$, $a_1\neq 0$, $a_2\neq 0$, and a $z\in\mathbb{Z}$ such that $a/n = w(zv/n)$ and 
$ c_{\text{WS}}\{L(n,v,\delta)\}^{-1} = \| a/n \| $. 
Without loss of generality assume $a_1>0$ and $a_2>0$. 
Because both entries of $v$ are coprime to $n$, there exists a $b_2 \in \mathbb{Z}$, $0<|b_2|\leq n/2$, and a $y \in \mathbb{Z}$ such that $(1,b_2)/n = w(y v/n)$. 
Because $\|a/n\| = c_{\text{WS}}\{L(n,v,\delta)\}^{-1} \leq \|(1,b_2)\|$, $|b_2|\geq a_2$ and $|b_2| \geq a_1$. 
From Lemma~1, $L(n,v,\delta) \cap \{ 1/(2n) \} \times [0,1]$ has one element. Let $(1/(2n),q)$ denote this element. 

Consider two cases on the sign of $b_2$. 
Firstly, when $b_2>0$. 
If $q<1-b_2/n$, we have $(1/(2n)+a_1/n,q+a_2/n) \in L(n,v,\delta)$ and $c_{\text{RS}}\{L(n,v,\delta)\}^{-1} \leq \|(1/(2n)+a_1,q+a_2/n) - (1/(2n),q)\| = \|a/n\| = c_{\text{WS}}\{L(n,v,\delta)\}^{-1}$. 
Otherwise, $(1/(2n)+1/n,q+b_2/n-1) \in L(n,v,\delta)$.
Because $a_1\leq n/2$ and $a_2\leq n/2$, $(1/(2n)+1/n+a_1/n,q+b_2/n-1+a_2/n) \in L(n,v,\delta)$ and $c_{\text{RS}}\{L(n,v,\delta)\}^{-1} \leq \|(1/(2n)+1/n+a_1/n,q+b_2/n-1+a_2/n) - (1/(2n)+1/n,q+b_2/n-1)\| = \|a/n\| = c_{\text{WS}}\{L(n,v,\delta)\}^{-1}$. 
This verifies that $c_{\text{RS}}\{L(n,v,\delta)\}^{-1} \leq c_{\text{WS}}\{L(n,v,\delta)\}^{-1}$ when $b_2>0$. 

Secondly, when $b_2<0$. 
If $q<1+b_2/n$, we have $(1/(2n)+a_1/n,q+a_2/n) \in L(n,v,\delta)$ and $c_{\text{RS}}\{L(n,v,\delta)\}^{-1} \leq \|(1/(2n)+a_1,q+a_2/n) - (1/(2n),q)\| = \|a/n\| = c_{\text{WS}}\{L(n,v,\delta)\}^{-1}$. 
Otherwise, because $|b_2|\leq n/2$, $(1/(2n)+1/n,q+b_2/n) \in L(n,v,\delta)$.
Because $a_1\leq n/2$ and $a_2\leq -b_2$, $(1/(2n)+1/n+a_1/n,q+b_2/n+a_2/n) \in L(n,v,\delta)$ and $c_{\text{RS}}\{L(n,v,\delta)\}^{-1} \leq \|(1/(2n)+1/n+a_1/n,q+b_2/n+a_2/n) - (1/(2n)+1/n,q+b_2/n)\| = \|a/n\| = c_{\text{WS}}\{L(n,v,\delta)\}^{-1}$. 
This verifies that $c_{\text{RS}}\{L(n,v,\delta)\}^{-1} \leq c_{\text{WS}}\{L(n,v,\delta)\}^{-1}$ when $b_2<0$. 

Combining both cases on the sign of $b_2$, $c_{\text{RS}}\{L(n,v,\delta)\}^{-1} \leq c_{\text{WS}}\{L(n,v,\delta)\}^{-1}$. 
Also because $c_{\text{RS}}\{L(n,v,\delta)\}^{-1} \geq c_{\text{WS}}\{L(n,v,\delta)\}^{-1}$, $c_{\text{RS}}\{L(n,v,\delta)\}^{-1} = c_{\text{WS}}\{L(n,v,\delta)\}^{-1}$.
\end{proof}

\begin{proof}[Proof of Theorem~3:]
Let $d = (b-ya)/z$. 
Then $b = y a + z d$, $a \cdot d=0$, and $\|d\|=\|a\|$. 
Because the $x_i + b$ is the point closet to the $x_i$ which does not lie on the line that passes through the $x_i$ and the $x_i+a$, $-1/2 \leq y \leq 1/2$ and $y^2 + z^2 \geq 1$. 

Consider two cases on the sign of $y$. Firstly, when $y\geq 0$. 
Then for each point $x_i \in L(n,v,\delta)$, the region that is nearest to $x_i$ than other lattice points $\{ z + iv/n + \delta/n + 1_d/(2n) : z \in \mathbb{Z}^p, i \in \mathbb{Z} \}$, i.e., \[ \| u-x_i \| = \min_{v \in \{ z + iv/n + \delta/n +1_d/(2n) : z \in \mathbb{Z}^p, i \in \mathbb{Z} \}}\| u-v \|,\] is the hexgon whose vertexes are $x_i+(1/2)a+\{(z^2-y+y^2)/(2z)\}d$, $x_i+(-1/2+y)a+\{(z^2+y-y^2)/(2z)\}d$, $x_i-(1/2)a+\{(z^2-y+y^2)/(2z)\}d$, $x_i-(1/2)a-\{(z^2-y+y^2)/(2z)\}d$, $x_i-(-1/2+y)a-\{(z^2+y-y^2)/(2z)\}d$, and $x_i+(1/2)a-\{(z^2-y+y^2)/(2z)\}d$. 
Therefore, $c_{\text{WS}}\{L(n,v,\delta)\} = [\{z^2+(z^2-y+y^2)^2\}^{1/2}/(2z)] \|a\|$. 
Similarly, when $y<0$, we have $c_{\text{WS}}\{L(n,v,\delta)\} = [\{z^2+(z^2+y+y^2)^2\}^{1/2}/(2z)] \|a\|$. 
Combining the two cases, we have $c_{\text{WS}}\{L(n,v,\delta)\} = [\{z^2+(z^2-|y|+y^2)^2\}^{1/2}/(2z)] \|a\|$. 
\end{proof}

\begin{proof}[Proof of Theorem~4:]
We have $L_j(n,v,\delta) = \{ z + isv/n + jv/n + \delta/n + 1_d/(2n) : z \in \mathbb{Z}^p, i \in \mathbb{Z} \} \cap [0,1]^d 
= \{ z + iv/(n/s) + \{\delta/s-1/2+1/(2s)+jv/s\}/(n/s) + 1_d/(2n/s) : z \in \mathbb{Z}^p, i \in \mathbb{Z} \} \cap [0,1]^d 
= L\{n/s,v,\delta/s-1/2+1/(2s)+jv/s\}$ for any $j$. 
Because all of the $L_j(n,v,\delta)$ are LLHDs in \eqref{eqn:LLHD} with the same generating vector $v$, they have the same WS, WA, WP, WD, WS2, and WF2 values. 
\end{proof}

\begin{proof}[Proof of Theorem~5:]
For any $zm/n + iv/n + \delta/n + 1_d/(2n)$ that lies in the $\prod_{k=1}^d [l_k,l_k+m/n]$, 
we have $n/m \{ zm/n + iv/n + \delta/n + 1_d/(2n) -l \} \in [0,1]^d$. 
Therefore, $zm/n + iv/n + \delta/n + 1_d/(2n) = r[ n/m \{ iv/n + \delta/n + 1_d/(2n)-l\} ] m/n + l = r\{ iv/m + (\delta-nl)/m + 1_d/(2m) \} (m/n) + l$. 
From Lemma~1, $zm/n + iv/n + \delta/n + 1_d/(2n) \subset L(m,v,\delta-ln) (m/n) + l$. 
Therefore, $R(n,m,v,\delta) \cap \prod_{k=1}^d [l_k,l_k+m/n] \subset L(m,v,\delta-ln) (m/n) + l$. 

On the other hand, for any $r\{ iv/m + (\delta-nl)/m + 1_d/(2m) \} \in L(m,v,\delta-ln)$, there exists a $z \in \mathbb{Z}^d$ such that 
$z + iv/m + (\delta-nl)/m + 1_d/(2m) \in [0,1]^d$. 
Therefore, $ \{ z + iv/m + (\delta-nl)/m + 1_d/(2m) \} (m/n) + l =  zm/n + iv/n + \delta/n + 1_d/(2n) \in \{ zm/n + iv/n + \delta/n + 1_d/(2n) : z \in \mathbb{Z}^p, i \in \mathbb{Z} \}$. 
Therefore, $L(m,v,\delta-ln) (m/n) + l \subset R(n,m,v,\delta)$. 
Also because $L(m,v,\delta-ln) (m/n) + l \subset \prod_{k=1}^d [l_k,l_k+m/n]$, 
$R(n,m,v,\delta) \cap \prod_{k=1}^d [l_k,l_k+m/n] =  L(m,v,\delta-ln) (m/n) + l$. 
\end{proof}

\begin{proof}[Proof of Theorem~6:]
Clearly, the $R(n,m,v,\delta-ln) = R(n,m,v,\delta-\tilde ln)$ if and only if $l-\tilde l \in \dot L(m,v) $. 
Therefore, $R(n,m,v,\delta) \cap \prod_{k=1}^d [l_k,l_k+q/n] = R(n,m,v,\delta-ln) \cap \prod_{k=1}^d [0,q/n] + l = R(n,m,v,\delta-\tilde ln) \cap \prod_{k=1}^d [0,q/n] + l 
= R(n,m,v,\delta) \cap \prod_{k=1}^d [\tilde l_k,\tilde l_k + q/n] + l - \tilde l$.  
\end{proof}

\begin{proof}[Proof of Theorem~7:]
Firstly, for any $v$ and $\tilde \delta$, $R(nm,m,v,\tilde\delta) m \cap \prod_{k=1}^d [w_k, w_k+1) = \{ R(n,m, \allowbreak v,\tilde\delta-nw) \cap [0,1]^d \} + w$. 
Let $l$ denote the greatest common divisor of $n$ and $m$. 
Among the $R(n,m,v,\tilde\delta-nw) \cap [0,1]^d$ with $w \in \{0,\ldots,m-1\}^d$, 
for any $y \in \{0,l,\ldots,m-l\}^d$, there are $l^d$'s many of them that equals to $R(n,m,v,\tilde\delta-y) \cap [0,1]^d$. 
Because $R(nm,m,v,\tilde\delta) m \cap [0,m]^d$ has $n^dm$ points, the sum of number of points of $R(n,m,v,\tilde\delta-nw) \cap [0,1]^d$ for $w \in \{0,\ldots,m-1\}^d$ is $n^dm$. 
Therefore, for $\delta$ randomly generated from the uniform distribution in $\tilde\delta + \{0,l,\ldots,m-l\}^d$, 
the expected number of design points for $R(n,m,v,\delta)$ is $n^d m^{-d+1}$. 
Further letting $\tilde \delta$ to be randomly generated from the uniform distribution in $\{0,\ldots,l-1\}^d$, 
we reach the conclusion that for $\delta$ randomly generated from the uniform distribution in $\{0,\ldots,m-1\}^d$, 
the expected number of design points for $R(n,m,v,\delta)$ is $n^d m^{-d+1}$. 
\end{proof}